\documentclass[]{interact}

\usepackage{subfigure}%

\usepackage{natbib}%
\usepackage{url}
\bibpunct[, ]{(}{)}{;}{a}{}{,}%
\renewcommand\bibfont{\fontsize{10}{12}\selectfont}%

\theoremstyle{plain}%

\theoremstyle{definition}

\theoremstyle{remark}

\usepackage{amsmath}
\usepackage{amsfonts,amssymb}
\usepackage{bm}
\usepackage{cases}
\usepackage{subfigure} 
\usepackage{booktabs}
\usepackage{multirow}
\usepackage{listings}
\usepackage{algorithm}  
\usepackage{algpseudocode} 
\usepackage{marvosym}
\newcommand{\tabincell}[2]{\begin{tabular}{@{}#1@{}}#2\end{tabular}} 
\usepackage{graphicx} %
\usepackage{epsfig}
\usepackage{fancyhdr}
\usepackage{setspace}
\usepackage{helvet}
\usepackage{makecell}

\usepackage{etoolbox}
\makeatletter
\patchcmd{\@maketitle}{{\receivedfont{\bfseries ARTICLE HISTORY\\}\@received\par}}{}{}{\PackageError{arxiv-preparation}{Could not remove template article history}{}}
\makeatother

\begin{document}

\articletype{Author manuscript}

\title{KT-EGO: A Knowledge Transfer Assisted Efficient Global Optimization Algorithm for Solving High-Dimensional Expensive Black-Box Problems}

\author{
\name{Qineng Wang\textsuperscript{a}, Liming Song\textsuperscript{a}, Yun Chen\textsuperscript{b}, Guangjian Ma\textsuperscript{b}, Zhendong Guo\textsuperscript{a}\thanks{Zhendong Guo. Email:guozhendong@xjtu.edu.cn} and Jun Li\textsuperscript{a} 
}
\affil{\textsuperscript{a}Institute of Turbomachinery, Xi'an Jiaotong University,Xi'an,Shaan Xi, China; \textsuperscript{b}AVIC Shenyang Engine Design Institute, Shenyang, Liao Ning, China}
}

\maketitle

\begin{abstract}
Many engineering problems involve optimizing a high-dimensional expensive black-box (HEB) design space. To solve such problems efficiently, we propose a knowledge transfer assisted efficient global optimization (EGO) algorithm, labeled as KT-EGO, which extends the EGO algorithm for solving problems over higher dimensions (i.e., $d>20$). Specifically, the original design space is divided into several low-dimensional subset design spaces. More importantly, in order to extract information from the subset design spaces to accelerate the progress of full optimization, we propose a surrogate-based data fusion strategy in KT-EGO. And further, a searching strategy with an adaptive variable range is devised to enhance the exploitation of promising areas. To show the effectiveness of our proposed algorithm, it is compared against the state-of-the-art algorithms over 12 benchmark functions and a 28-dimensional engineering optimization for the design of compressor blade, which fully validates the effectiveness of the KT-EGO for solving HEB problems.
\end{abstract}


\noindent\textit{Abridged author version of the article published in Engineering Optimization 55(12), 2015--2033 (2023), DOI: 10.1080/0305215X.2022.2139374. Detailed supplemental tables are omitted; retained engineering results and variable numbering follow the published main article.}

\section{Introduction}\label{sec1}
\par
Many engineering problems, such as turbomachinery design ~\citep{he_multi-objective_2021,li_elite-driven_2022,song_research_2016}, involve optimizing an expensive black-box problem over high-dimensional design space~\citep{Karim_framework_2014,Wang_Committee_2017,baert_aerodynamic_2020,li_three-level_2022}.
Such problems are labeled as high-dimensional expensive black-box (HEB) problems~\citep{shan2010survey}, which are also known as large-scale black-box problems~\citep{peng2018multimodal}.
As pointed out by \citet{omidvar_review_2021-1}, the definition of the HEB problem may vary from field to field.
For instance, when solving the optimization problems that involved in aircraft and automotive design, the problem with dimensions larger than 10 can be treated as HEB problems~\citep{shan_metamodeling_2010}. 
Due to the huge cost of the simulation process, such HEB problems cannot be easily solved by the conventional surrogate-based optimization~\citep{regis_combining_2013,dong_surrogate-based_2018,bouhlel_gradient-enhanced_2019}(SBO) algorithms and genetic algorithms~\citep{melanie_introduction_1998}.
In this paper. we follow the definition in ~\citep{zhan_fast_2021}, which defines the HEB problem as follows. That is, (1) the problem dimension is greater than 20; 
and (2) only a few hundreds or thousands of evaluations can be affordable~\citep{haftka2016parallel}. 

To solve such HEB problems efficiently, some studies propose to make use of surrogate models to assist the optimization search of evolutionary algorithms, which are known as surrogate-assisted evolutionary algorithms~\citep{dong2020surrogate,dong2021surrogate,li_surrogate-assisted_2021,wang_surrogate-assisted_2022}.
Alternatively, some others propose to solve the HEB problem with more advanced surrogate-based optimization (SBO) algorithms.
In particular, one of the promising SBO algorithms for solving the HEB problem is Nash-EGO~\citep{xu_nash_2018,xu_study_2019}. 
In Nash-EGO, the original high-dimensional design space is first divided into several low-dimensional subset design spaces. 
Then, optimizations are carried out over these subset design spaces.
After that, the optimal solutions of subset design spaces are combined to obtain the best solution to the original HEB problem.
Following this way, Nash-EGO extends the efficient global optimization (EGO) algorithm~\citep{jones1998efficient,jones_taxonomy_1997}, which is well-known to be sample-efficient for solving problems over domains of less than 15 dimensions, for solving high-dimensional problems.
\par
While showing promise for solving the HEB problems, the performance of Nash-EGO can be further improved.
Specifically, the evaluated training samples of subset design spaces that were used in previous optimization cycles of Nash-EGO are discarded. 
However, the previously evaluated samples can  contain useful information to facilitate the optimization search in the following iterations.
Hence, inspired by the fact that we humans often gain experiences from the source tasks or dataset to help resolve the problem at hand, we propose a knowledge transfer~\citep{pan_survey_2010} assisted efficient global optimization algorithm (KT-EGO), which intelligently gains knowledge from the previously evaluated samples to accelerate the optimization progress.
\par
Furthermore, considering the fact that a more local search is desired when the algorithm reaches the neighborhood of the true optimal solution, a kriging-based search strategy with adaptive variable ranges is devised to enhance the exploitation of promising areas at the later subset optimization stage in the KT-EGO.
Moreover, for non-separable optimization problems~\citep{mahdavi_metaheuristics_2015,schaefer_large-scale_2010,yang_large_2008}, the variables in the design space can highly interact.
As a consequence, it is almost impossible to distinguish the variable interactions and therefore decompose the design space correctly in one shot.
Therefore, a random decomposition strategy is proposed to capture the variable interactions in the searching process of KT-EGO. 
With the above, the main contributions of this paper can be summarized as follows:
\par
(1) We propose a knowledge transfer assisted efficient global optimization algorithm, namely KT-EGO, to solve the HEB problems more efficiently.
In particular, after decomposing the original design space into subset design spaces, a surrogate-based data fusing strategy is proposed to enable knowledge transfer in KT-EGO, which draws useful information from the previously evaluated samples to boost the algorithm performance.
\par
(2) Considering the joint effects of variables in high-dimensional design space, a random decomposition strategy is proposed to extend the KT-EGO algorithm for solving non-separable HEB problems.
Moreover, a kriging based local exploitation strategy is devised to accelerate the optimization progress at the later stage of KT-EGO.
\par
(3) Through tests on 12 benchmark problems and an engineering problem of compressor blade, the effectiveness of our proposed KT-EGO has been well demonstrated.
\par
The remainder of this paper is organized as follows.
In Section 2, the background and related work of KT-EGO are introduced.
And then, the details of KT-EGO are illustrated in Section 3.
After that, the proposed KT-EGO algorithm is tested on benchmark functions and an engineering optimization problem in Section 4.
And finally, we draw conclusions in Section 5.
\section{Research Background}\label{sec2}
\subsection*{2.1 The Nash-EGO algorithm}
\begin{figure}[ht]
\begin{center}
\includegraphics[scale=0.55, trim = 0 0 0 0]{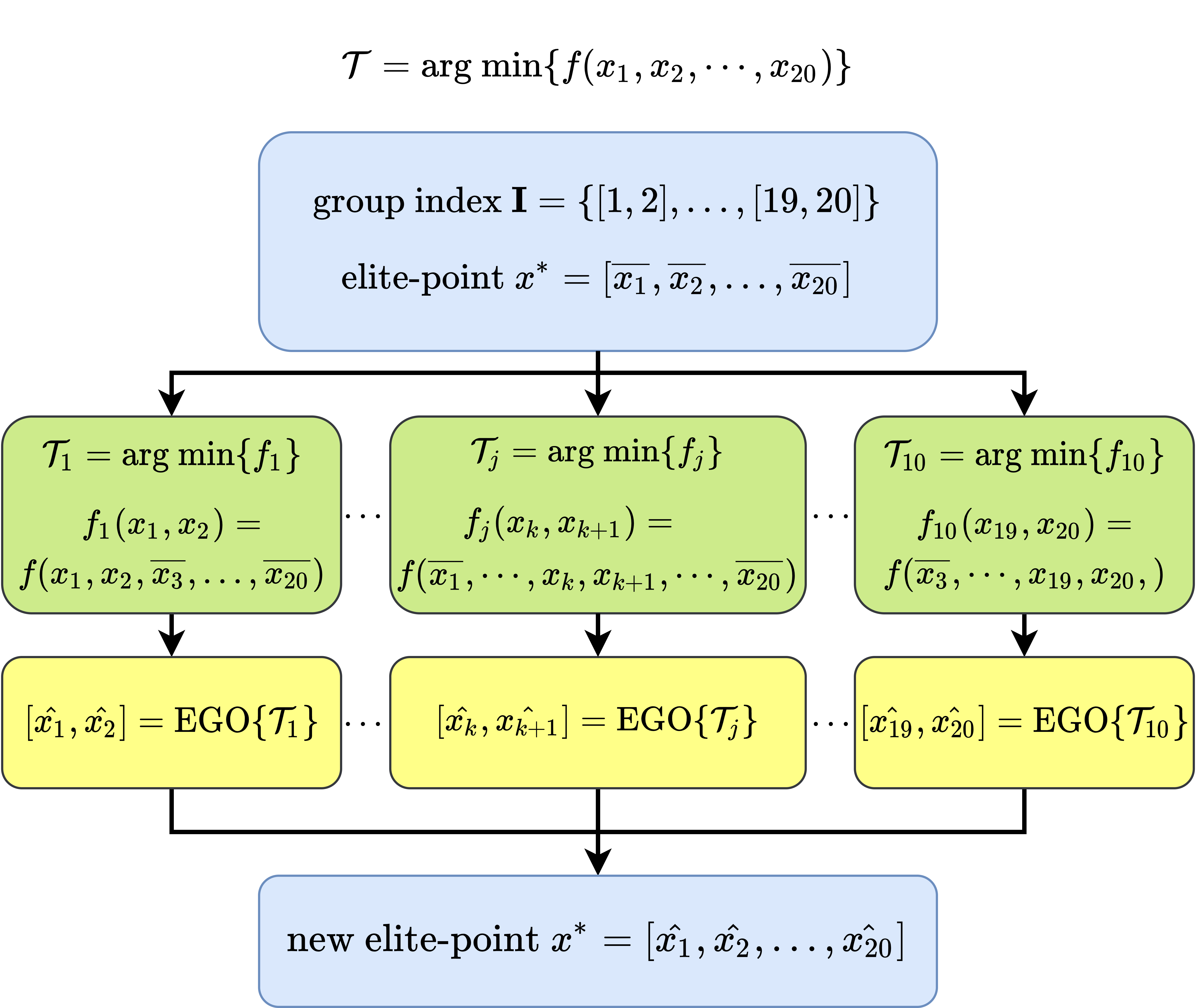}
\end{center}
\caption{Decomposition and combination strategy in one cycle of the Nash-EGO algorithm}   
\label{fig:kt-original-1}
\end{figure}

The Nash-EGO algorithm is a decomposition based algorithm, which starts the optimization search from an elite point.
Here, the elite point is defined as the current best solution at $c^{\text{th}}$ cycle, which is a random point at the first optimization cycle.
The complete searching process of Nash-EGO at one optimization cycle is shown in Figure~\ref{fig:kt-original-1}.
Firstly, by using the elite point as a reference point, the original high-dimensional design space is divided into several low-dimensional subset design spaces.
In particular, except for the variables in a subset design space, the remaining sample values associated with the subset design space are fixed as the same as the elite point.
Secondly, optimizations are carried out in each subset design space, by using the EGO algorithm as the optimizer.
Thirdly, the optimal solutions of the subset design spaces are combined to update the elite point.
The Nash-EGO will repeat the above process until the termination condition is met.

As mentioned earlier in the Introduction part, by dividing the original high-dimensional design space into low-dimensional subset design spaces, the Nash-EGO algorithm makes it possible to extend the sample-efficient EGO algorithm for solving HEB problems.
However, in each iteration loop of Nash-EGO, the design of the experiment (DoE) should be carried out for each subset design space. 
In other words, the training samples of a subset design space in a previous optimization cycle are note reused to facilitate optimization search in the following iterations.
Nonetheless, the training samples of the previous iterations may 
contain useful information to accelerate the optimization process of the current optimization cycle.
Hence, we propose KT-EGO, which extracts useful information from all the previous training samples to promote the optimization search in subset design spaces.

\subsection*{2.2 Transfer Optimization and Multi-Form Optimization}
The basic idea of knowledge transfer is to intelligently gain knowledge from the related source data or tasks to achieve better solutions with even fewer computational costs~\citep{pan_survey_2010,guo_parallel_2021}, which is an emerging topic known as transfer learning in the machine learning community.
The idea of knowledge transfer has also drawn wide attention for solving black-box optimization problems, known as transfer optimization (TO), and a series of success stories has been reported in this relatively young field.
According to~\citep{gupta_insights_2018,tan_evolutionary_2021,guo_generative_2022}, TO can be classified into three categories, i.e., sequential transfer optimization (STO), multitasking optimization (MTO), and multi-form optimization (MFoO).
While STO and MTO deal with distinct optimization tasks that are presumed to be related, MFoO exploits alternate formulations of a single target task.
More specifically, in the hope to improve the efficiency of a search algorithm, MFoO proposes joint optimization of several alternate formulations, with continuous knowledge transfer between them:
\begin{equation}
\begin{array}{c}
T = \mathop {\min }\limits_{{\bf{x}} \in {\rm{{\cal X}}}} f({\bf{x}})\\
Q_{c}^{\text{MFoO}}   (T\| T_{1},\cdots,T_{m},K(c)) - Q_{c}(T) \ge 0
\end{array}
\end{equation}
where, $T$ is the target optimization problem, and $Q_c$ is a measure quantifying the quality of solution(s) obtained at the $c^{\text{th}}$ cycle, ${T_1}, \cdots ,{T_m}$ are alternate formulations of $T$, and $K(t)$ is the strategy to facilitate knowledge transfer between the alternate formulations.
In our study, $T_m$ is the optimization process that is carried out over a subset design space, and we propose a surrogate-based data fusion strategy as $K(c)$ to facilitate knowledge transfer in KT-EGO, which will be discussed in detail in the next section. 
\subsection*{2.3 Polynomial Chaos Expansion}

Polynomial chaos expansion (PCE)~\citep{wiener_homogeneous_1938, sudret_global_2008} is a popularly used surrogate technique for representing a random variable in terms of a polynomial function of other random variables~\citep{gupta_insights_2018,tan_evolutionary_2021}.
It can be regarded as a fitting regression model surrogate model when these random variables are uniformly distributed\citep{palar_polynomial-chaos-kriging-assisted_2017,palar_efficient_2018}.   
As it has good approximation accuracy over high-dimensional design space, PCE is used to build a surrogate-based data fusion strategy in this article.
\par 
Formally, the function prediction of PCE can be expressed as:
\begin{equation}
\hat{y}_\text{PCE}(x) \approx {{\cal M}^{(PCE)}}(x) = 
\sum\limits_{{\bf{\alpha }} \in {\cal A}} {a_{\bf{\alpha }}} {\psi _{\bf{\alpha }}}(x)
\end{equation}
where, ${\psi _{\bf{\alpha }}}(x)$ is a sequence of polynomials; 
$\alpha$ is the multi-index of the multivariate polynomial ${\psi _{\bf{\alpha }}}(x)$,
${\bf{\alpha }} = \{ {{\alpha _1}, \cdots ,{\alpha _D}} \}$;
$D$ is the dimension of input variables.
Multivariate polynomial ${\psi _{\bf{\alpha }}}(x)$ is the product of multiple orthogonal single-variable polynomials $\psi _{{\alpha _i}}^{(i)}\left( {{X_i}} \right)$, 
as ${\psi _{\bf{\alpha }}}(x) = \prod\limits_{i = 1}^D {\psi _{{\alpha _i}}^{(i)}} (x_i)$.
\par 
In our study, the PCE will be built with uniformly distributed variables. Correspondingly, the Legendre polynomial is selected as the orthogonal base when building PCE.
\begin{equation}
{\left\langle {\psi _i^{(k)},\psi _j^{(k)}} \right\rangle _k} = \int_{{{\cal D}_k}} {{\psi _i}} (x){\psi _j}(x){f_{{X_k}}}(x)dx = {\delta _{ij}}
\end{equation}
where, and ${\delta _{ij}}$ is 1 if $i = j$ and 0 if $i \neq j$. More details on PCE can refer to~\citep{xiu_numerical_2010,sudret_global_2008}..

\subsection*{2.4 Kriging and Hierarchical Kriging}
\subsubsection*{2.4.1 Kriging}
\par
Kriging is a popular surrogate technique~\citep{Krige_statistical_1952}.
The kriging prediction $Y_{KG}$ at unknown site $x$ is built as a trend function $f(x)$ plus a normal random process $Z(x)$ as:
\begin{equation}
{Y_{KG}}({\bf{x}}) = f({\bf{x}}) + Z({\bf{x}})
\end{equation}
where, $f(x)$ is usually a constant, linear or quadratic polynomial, and the constant is most widely used; $Z(x)$ describes the local features of $Y$ around the $n$ sample points $X = \{ x^{(1)}, \cdots, x^{(n)}\}$, which has zero mean and a co-variance function as:
\begin{small}
\begin{equation}
{\mathop{\rm cov}} \left[ {Z({\bf{x}}),Z({{\bf{x}}^{(i)}})} \right] = {\sigma ^2}\exp \left( { - \sum\limits_{h = 1}^d {{\theta _h}{{\left\| {{x_h} - x_h^{(i)}} \right\|}^2}} } \right)
\end{equation}
\end{small}
The function prediction and related mean squared error (MSE) at an unknown point $\bf{x}$ can be expressed as:
\begin{small}
\begin{equation}
\begin{split}
&{{\hat y}_{KG}}({\bf{x}}) = \hat \mu  + {{\bf{r}}^T}{{\bf{R}}^{ - 1}}({\bf{y}} - {\bf{1}}\hat \mu )\\
&{s_{KG}}({\bf{x}}) = 
{\sigma ^2} \{1 - {\bf{r}}^T ({\bf{x}}) {\bf{R}}^{-1} {\bf{r}}({\bf{x}}) + \\
&( 1 - {\bf{1}}^T  {\bf{R}}^{ -1} {\bf{r}}^T ({\bf{x}}))
{{( {{{\bf{1}}^T}{{\bf{R}}^{-1}}{\bf{1}}})}^{ - 1}}
( 1 - {\bf{l}}^T  {\bf{R}}^{-1}  {\bf{r}}^T ({\bf{x}}) )^T \}\\
\end{split}
\end{equation}
\end{small}
where $\mu$ is the regression constant as 
$\mu=\left(
\mathbf{1}^{T} \mathbf{R}^{-1} \mathbf{1}
\right)^{-1} 
\mathbf{1}^{T} \mathbf{R}^{-1} \mathbf{y}_{S}$;
$\mathbf{R}$ is the correlation matrix $\mathbf{R}:=\left(R\left(\mathbf{x}^{(i)}, \mathbf{x}^{(j)}\right)\right)_{i, j} \in \mathbb{R}^{n \times n}$, and $\mathbf{r}$ is the correlation vector
$\mathbf{r}:=\left(R\left(\mathbf{x}^{(i)}, \mathbf{x}\right)\right)_{i} \in \mathbb{R}^{n}$.
\subsubsection*{2.4.2 Hierarchical Kriging}
\par 
Hierarchical kriging (HK)~\citep{han_hierarchical_2012} is a popularly used multi-fidelity surrogate technique, which combines both high- and low-fidelity samples to build approximation models~\citep{qian_bayesian_2008,zhang_variable-fidelity_2018}.
Assuming that the high-fidelity function can be modeled as a scaled low-fidelity function $f_{L}(\bf{x})$ plus a Gaussian process ${Z}(\bf{x})$, the hierarchical kriging (HK) can be formulated as:
\begin{equation}
Y_{HK}({\bf{x}}) = {\beta _0} f_{L}(\bf{x}) + \bf{Z}(x)
\end{equation}
Given the high-fidelity samples $\{ \bf{X},\bf{y}\}$ and the low-fidelity model $f_{L}(\bf{x})$, the function prediction and MSE at an unknown point $\bf{x}$ can be expressed as:, 
\begin{small}
\begin{equation} 
\begin{split}
&\hat y_{HK}({\bf{x}}) = {\beta _0}{f_L}({\bf{x}}) + {r^T}({\bf{x}}){R^{ - 1}}({\bf{y}} - {\beta _0}F)\\
&{s^2_{HK}}({\bf{x}}) = 
{\sigma^2} ( 1 - {r^T} {R^{ - 1}}r + ( {{r^T}{R^{ - 1}}F - {f_L}({\bf{x}})}){{( {{F^T}{R^{ - 1}}F} )}^{ - 1}}{{( {{r^T}{R^{ - 1}}F - {f_L}({\bf{x}})})}^T})
\end{split}
\end{equation} 
\end{small}
where, $\beta_{0}$ is the scaling factor modeling the correlations between the high- and low-fidelity model, and $\beta _0= (\bf{F}^{T} \bf{R}^{-1} \bf{F})^{ - 1} (\bf{F}^{T}\bf{R}^{ - 1}Y)$;
and $\bf{F}$ is the low-fidelity function vector of $X$, i.e.,
${\bf{F}} = [f_\text{L}(x^{(1)}), \cdots ,f_\text{L}(x^{(n)}) ]^T$;
and $\bf{R}$ and $\bf{r}$ are the correlation matrix and correlation vector of the related multi-fidelity Gaussian process.
More details on HK can refer to~\citep{han_hierarchical_2012}.

\section{Proposed Method}\label{sec3}
\par
Before a detailed discussion of our proposed KT-EGO, we show the motivations behind KT-EGO as follows.
Firstly, considering the joint effects of design variables, it may not be easy to divide the original problem into subset design spaces that best fit the target landscape in one shot. 
Hence, instead of dividing the original design space by the sequence of the variable index as Nash-EGO, we propose a random decomposition approach, which updates the subset design spaces in every optimization cycle.
Secondly, 
inspired by the fact that we humans often gain experiences from the source tasks or dataset to help resolve the problem at hand, we propose a surrogate-based data fusion strategy to facilitate the optimization search.
More specifically, as capturing the global trend of the target problem can help the algorithm to arrive at the neighborhood of the true optimal more quickly~\citep{sasena2002exploration,ponweiser2008clustered}, the proposed data fusion strategy extracts trend information from the previously evaluated samples to facilitate the optimization search in subset design space, particularly that at the beginning of the optimization process.
Thirdly, 
at the later stage of each subset optimization process, especially when the algorithm arrives at the vicinity of the true optimal, a more local search of promising areas is desired.
Moreover, the function trend in the local promising areas can be different from the global trend of the target problem. Thereby, the knowledge gained from the global surrogate models may harm the local exploitation process~\citep{wang_transfer_2020,guo_analysis_2018}.
Hence, we propose an adaptive local exploitation strategy at the later stage of the subset optimization process.
\par
\begin{figure}[htp]
\begin{center}
\includegraphics[scale=0.85, trim = 0 0 0 0]{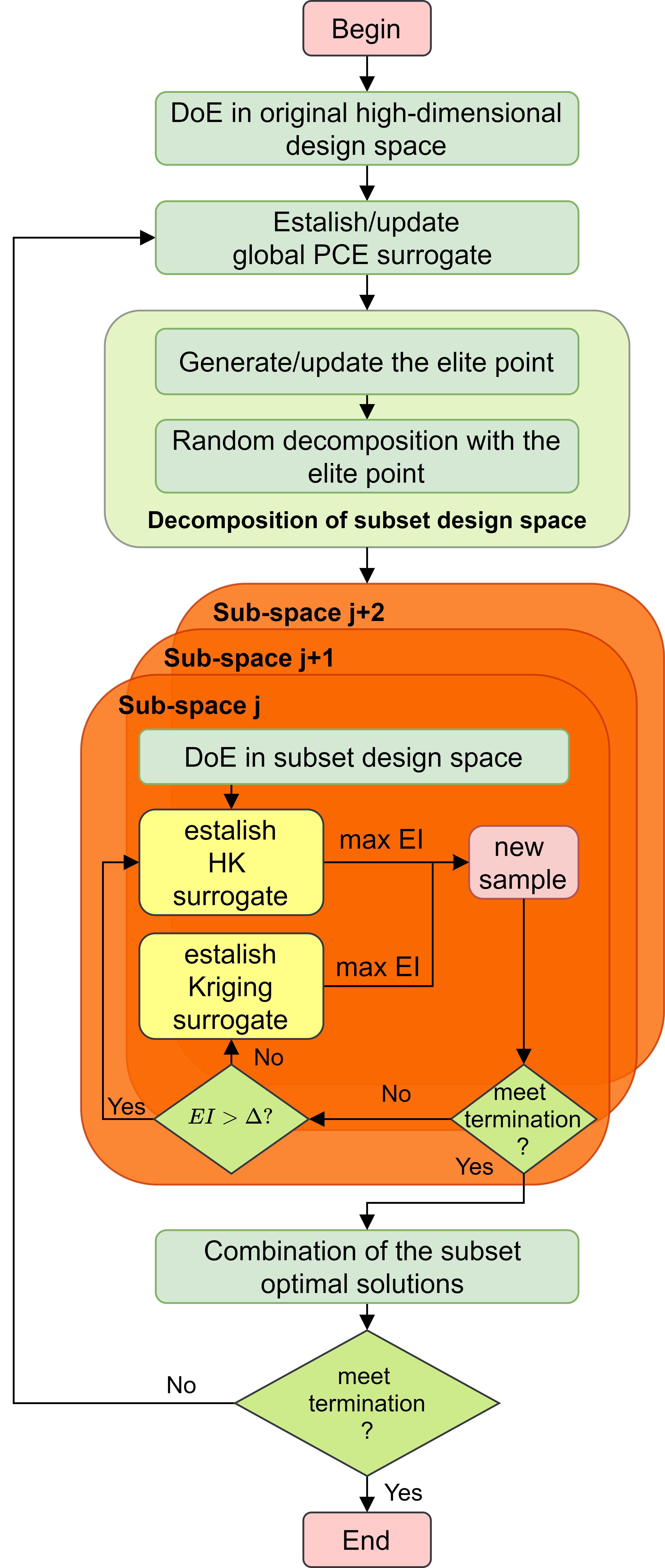}
\end{center}
\caption{The framework of the KT-EGO algorithm}   
\label{fig:kt-original-2}
\end{figure}

With the above in mind, we present the framework of proposed KT-EGO algorithm in Figure~\ref{fig:kt-original-2}, which consists of four parts, i.e., (1) Design of Experiment (DoE) in the original high-dimensional space; (2) Build or update the surrogate over the original high-dimensional space; (3) Decomposition of the high-dimensional space into subset spaces with elite points and (4) Optimizations in subset design spaces.  

Specifically, the Latin hypercube sampling (LHS) is used for the DoE of the high-dimensional space in (1).
Then, a PCE surrogate is built with all the training samples, which tries to capture the global trend of the original high-dimensional design space in (2). 
As (1) and (2) are regular processes like most surrogate-based optimization algorithms, we would be focused on illustrating (3) and (4) in the following paragraphs.

\subsection*{3.1 Decomposition of subset design spaces}

Recalling Figure~\ref{fig:kt-original-1}, an elite point should be first generated before dividing the original high-dimensional design space into subset spaces. 
Similar to the Nash-EGO algorithm shown in Section 2.1, the elite point is selected as the sample with current best objective function value, labeled as ${\bf{x}}^{*}$.
In the meantime, different from Nash-EGO, we propose a random decomposition strategy to divide the original design space into subset spaces, which randomly and dynamically changes the division of subset spaces at each optimization cycle.

More specifically, supposing the original high-dimensional design space can be divided into $r$ subset design spaces, the indexes of these subset design spaces are denoted by ${I_1},{I_2} \cdots {I_r}$, as shown in Eq.(9).
In the Nash-EGO algorithm, the subset design spaces are divided directly according to the sequence of the variable indexes of original problem.
However, considering the fact that the variable interactions of original high-dimensional design space are unknown, and it may not be that easy to divide the original design spaces into subset spaces that best fit the original landscape in one shot. Hence, we propose to reorder the indexes of variables randomly, and accordingly obtain the subset spaces with the new ordered variable index.
Such random process will be repeated in each iteration cycle by using the Matlab function "randperm".

For the simplicity of illustration purpose, three variables are assigned to each subset design space in the case shown in Eq.(9).
Nevertheless, we can assign a different number of variables in each subset design space in KT-EGO.
\begin{small}
\begin{equation}
\begin{array}{l}
I^{\text{Nash-EGO}} = \{ \underbrace {\{ 1,2,3\} }_{{I_1}},\underbrace {\{ 4,5,6\} }_{{I_2}},...,\underbrace {\{ D - 2,D - 1,D\} }_{{I_r}}\} \\
I^{\text{KT-EGO}} = \{ \underbrace {\{ {l_1},{l_2},{l_3}\} }_{{I_1}},\underbrace {\{ {l_4},{l_5},{l_6}\} }_{{I_2}},...,\underbrace {\{ {l_{D - 2}},{l_{D - 1}},{l_D}\} }_{{I_r}}\} \\
{\{ {l_1},{l_2}, \cdots ,{l_D}\}} = randperm(D)
\end{array}
\end{equation}
\end{small}

\subsection*{3.2 Optimization in subset design space}

In this section, the DoE in subset design space and two strategies to promote the optimization search in subset spaces are discussed in detail. 

\subsubsection*{3.2.1 DoE in subset design spaces}
After finishing the decomposition of subset design space, LHS is used to generate the samples in each subset space, which is labeled as ``DoE in subset design space'' in Figure~\ref{fig:kt-original-2}.
Take $I_1$ of $I^{KT-EGO}$ in Eq.(9) (see the equation in Section 3.2) for example, when carrying out DoE in subset space $I_1$, only the sample value with the indexes $l_1, l_2$ and $l_3$ will be changed using LHS, the remaining variable values of a sample in $I_1$ will be fixed exactly the same as the elite point ${\bf{x}}^{*}$. 

Note that, at each optimization cycle of KT-EGO, DoE will be carried out for each subset space in a parallel way. 
Then, a number of iterations will be conducted for the optimizations of subset design spaces simultaneously. 
The iteration setting used in the benchmark comparisons is specified in Section 4.1.

\subsubsection*{3.2.2 knowledge transfer assisted optimization}

Recalling the iteration process in Nash-EGO shown in Figure~\ref{fig:kt-original-1}, the samples of a subset design space cannot be directly used in another subset design space.
Hence, the optimizations of subset design spaces of Nash-EGO are carried out from scratch in each optimization cycle, which does not draw any information from the previously evaluated samples to facilitate the current optimization search.
Differently, we propose a surrogate-based data fusion strategy to gain knowledge from the previously evaluated samples, which enables knowledge transfer to accelerate the optimization progress of each subset design space, particularly at the beginning of the optimization.
\par
More specifically, though the samples of a subset design space cannot be directly used in another subset design space, the sample information of all subset design spaces can be encoded in a surrogate of the original high-dimensional design space~\citep{shan_development_2009,rennen_subset_2009}.
Thereby, the function values at any site of the original and subset design space can be predicted. 
Furthermore, the predictions of the global surrogate using PCE are used as low-fidelity samples to augment the samples in each subset design space, which helps to achieve better trend predictions in subset design space and therefore boosts the optimization progress.
\par
With the above, the surrogate-based data fusion strategy and related subset design space optimization process are shown in Figure~\ref{fig:kt-original-3}, which can be described as follows:
\begin{figure*}[ht]
\begin{center}
\includegraphics[scale=0.65, trim = 0 0 0 0]{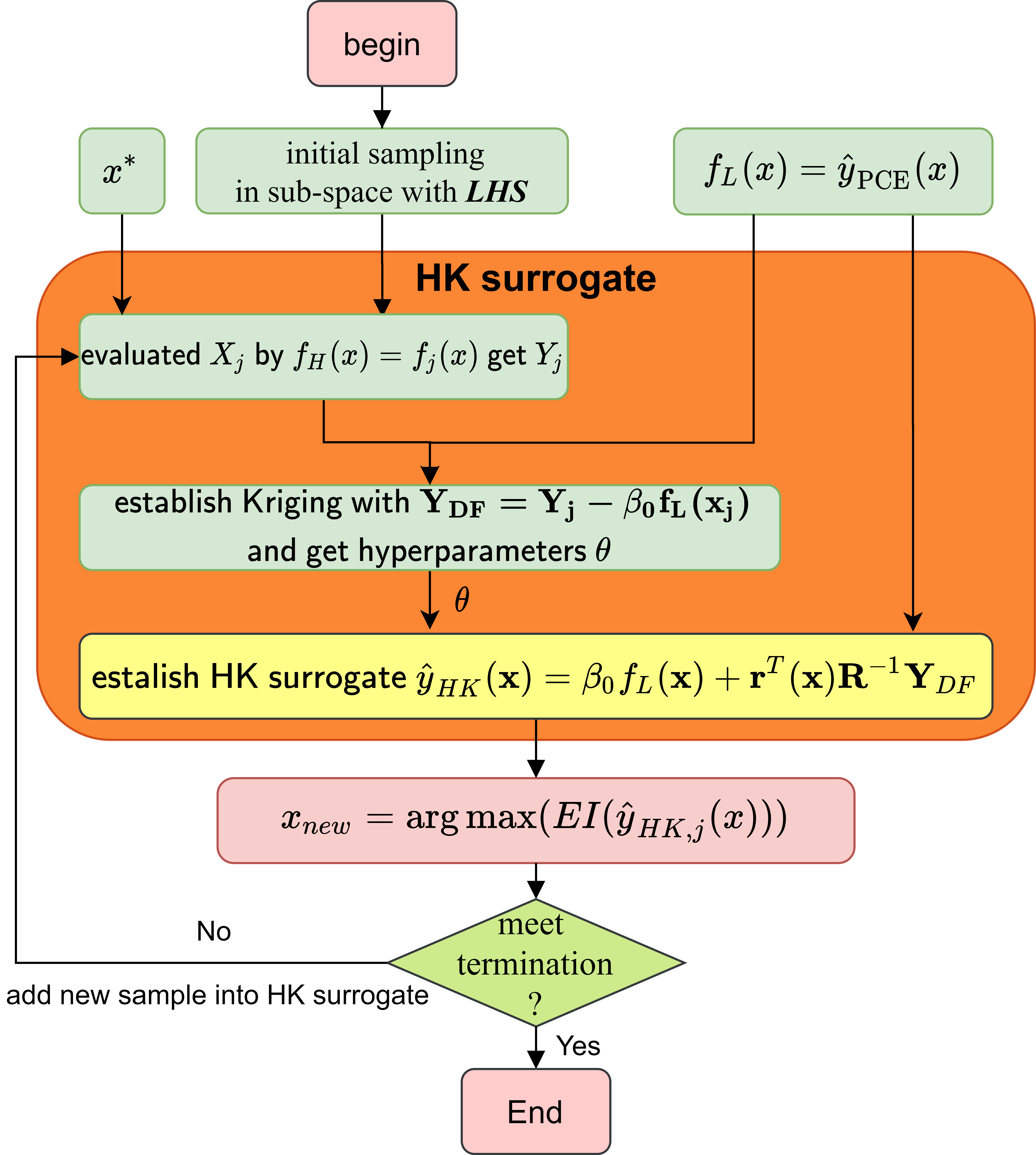}
\end{center}
\caption{Surrogate-based data fusion strategy and related subset design space optimization process}   
\label{fig:kt-original-3}
\end{figure*}
\par
(1) The PCE surrogate, which is built with all the training samples of the original high-dimensional space is used as low-fidelity surrogate to help the multi-fidelity surrogate modeling of an subset design space. The surrogate of the original high-dimensional space is labeled as ${\hat y_{\text{PCE}}}$;
\par
(2) For the optimization in a subset design space, the LHS is used to obtain the initial training set with $d$ dimension containing $2d$ samples, which are evaluated by the simulation methods such as CFD.
It is worth noting that, $d$ is the dimensions of the current sub-space, which is different from the dimension of the original space $D$. 
So, a $d$ dimension location can be transferred into a $D$ dimension location by
overwriting the $d$ corresponding variables in $x^{*}$ respectively.
In the meantime, the elite point $x^{*}$, i.e. the current best solution of the $j^{\text{th}}$ optimization cycle is also added to the training sample set.
\par
(3) The training samples are also evaluated by ${\hat y_{\text{PCE}}}$, which are treated as low-fidelity samples.
Then, combining the low-fidelity samples with the training samples of the subset design space, a multi-fidelity surrogate is built for the subset design space by HK.
\par
(4) The acquisition function of expected improvement (EI) is used to find the promising samples to query, which is formulated as~\citep{jin_comparative_2000,liu_survey_2018,guo_calibrated_2021}:
\begin{small}
\begin{equation}
\begin{split}
E{I_{KT}}({{\bf{x}}_{sub,k}}) &= ( {{f_{\min }} - {{\hat y}_{HK}}({{\bf{x}}_{sub,k}})} )\Phi (u) + s_{HK}({{\bf{x}}_{sub,i}})\phi (u)\\
u &= {{( {{f_{\min }} - {{\hat y}_{HK}}({{\bf{x}}_{sub,k}})} )} \mathord{/
 {\vphantom {{( {{f_{\min }} - {{\hat y}_{HK}}({{\bf{x}}_{sub,k}})} )} {s({{\bf{x}}_{sub,k}})}}}} {s_{HK}({{\bf{x}}_{sub,k}})}}
\end{split}
\end{equation}
\end{small}
where, the subscript KT is short for knowledge transfer, ${\bf{x}}_{sub,k}$ is the sample site in the $k^{\text{th}}$ subset design space, and $f_{min}$ is the current best solution in subset design space.
The expressions of $\hat y_{\text{HK}}$ and $s_{\text{HK}}$ can refer to Eq.(8), and $\Phi \left(\cdot\right)$ and $\phi \left(\cdot\right)$ denote the standard normal distribution and density probability functions. 
\par
(5) Evaluate the new samples suggested by Eq.(10), and update the HK surrogate. 
\par
(6) Repeat the process in Steps (3) to (5) until the termination condition is met.
\par
Also note that the PCE model built in Step (1) will be reused for the optimizations of all subset design space in this cycle.
 
Instead, it can be much easier to capture the overall trend of the high-dimensional landscape. 
Hence, the PCE model with low-order polynomials instead of interpolation technical like kriging is used to build the surrogate of the high-dimensional design space.

By following this way, the trend information of the high-dimensional design space can be further encoded into the subset design space by using HK. 
Then, by using the multi-fidelity surrogate $\hat y_{HK}$ in Eq.(8) when selecting new samples to query, the knowledge of previously evaluated samples is encoded to facilitate the optimization search in a subset design space.
\par
After all the sub-spaces have been optimized,
a new combined sample is constructed by concatenating the optimal value in  corresponding to each variable.
Then this new combined sample will be evaluated by performance evaluation method such as CFD.

\subsubsection*{3.2.3 Local exploitation with adaptive variable range}
\par
\par

\begin{figure}[ht]
\begin{center}
\includegraphics[scale=0.65, trim = 0 0 0 0]{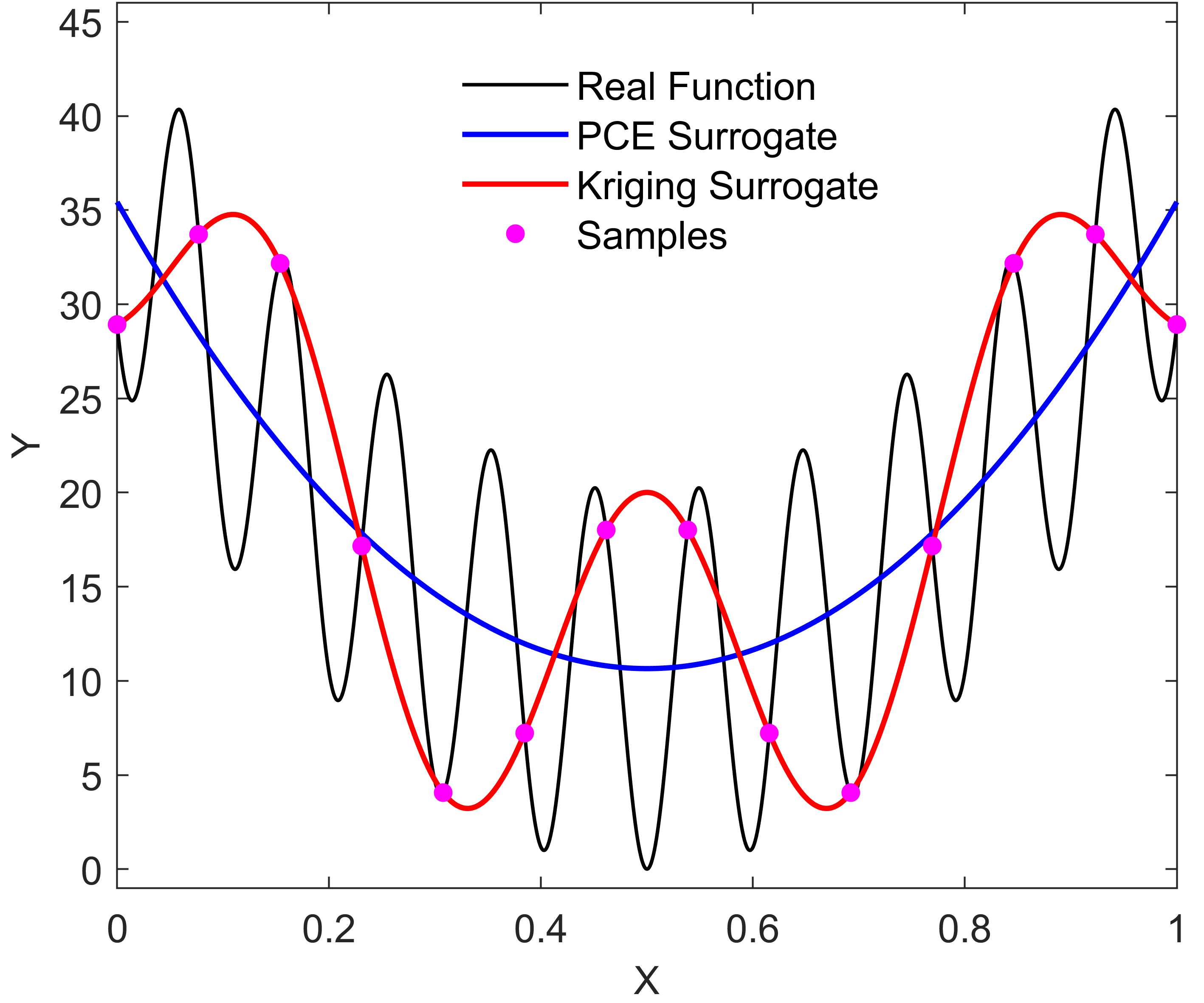}
\end{center}
\caption{Comparison of PCE with low-order polynomials and kriging surrogate}   
\label{fig:kt-original-4}
\end{figure}
As noted in the first paragraph of this section, at the later stage of the subset optimization process, particularly when the algorithm arriving at the neighborhood of the true optimal, an enhanced local search of promising areas is desired.
More importantly, as the function trend in promising areas can be different from the global trend of the target problem, the global trend information may mislead the optimization search in the subset design space at this stage.
Hence, we split the optimization process of subset design space into two stages.
In the first stage, the HK is used in the EI formulation to guide the optimization process, as illustrated with Eq.(10).
Then, at the second stage of the subset optimization process, the kriging surrogate instead of HK is used to build an approximation model in the related subset design space.
In the meantime, the two stages will be split by setting a threshold value of EI, labeled as $\Delta$.
And accordingly, the related formulations are shown as below:
\begin{small}
\begin{equation}
\begin{array}{c}
EI({{\bf{x}}_{sub,j}}) = \left( {{f_{\min }} - {\hat y}({{\bf{x}}_{sub,j}})} \right)\Phi (u) + s({{\bf{x}}_{sub,j}})\phi (u)\\
u = {{\left( {{f_{\min }} - {{\hat y}_{}}({{\bf{x}}_{sub,j}})} \right)} \mathord{\left/
 {\vphantom {{\left( {{f_{\min }} - {{\hat y}_{}}({{\bf{x}}_{sub,j}})} \right)} {s({{\bf{x}}_{sub,j}})}}} \right.
 \kern-\nulldelimiterspace} {s_{}({{\bf{x}}_{sub,j}})}}
\end{array}
\end{equation}
\end{small}

\begin{equation}
\begin{small}
\begin{gathered}
\hat{y}= \begin{cases}\hat{y}_{H K} & \text { if } E I_{\max }>\Delta \\
\hat{y}_{K G} & \text { if } E I_{\max } \leq \Delta\end{cases} \\
s^{2}= \begin{cases}s_{H K}^{2} & \text { if } E I_{\max }>\Delta \\
s_{K G}^{2} & \text { if } E I_{\max } \leq \Delta\end{cases}
\end{gathered}
\end{small}
\end{equation}
where, ${\bf{x}}_{sub, k}$ denotes the sample point of the $k^{th}$ subset design space, and $EI_{\max }$ in Eqs. (12) denote the maximum EI value in the $k^{th}$ subset design space. 
By following the suggestions in~\citep{jones1998efficient}, the value of $\Delta$ is set as $10^{-3}$ in this paper.
The expressions of $\hat y$ and $s^2$ can refer to Eqs. (6) and (8).

More specifically, when $EI_{\max } \leq \Delta$, we use the space reduction method that proposed in~\citep{regis_trust_2016,dong_hybrid_2018} to adaptively change the searching range of the $k^{th}$ subset space for enhanced local exploitation. 
The detailed process is as follows:

(1) Generate $N$ samples over the $j^{th}$ subset design space and approximate the related function value using the HK surrogate (see Eq.(7));

(2) Gather the selected local candidate points and their predicted function values into a matrix $D_j$, as follows:
\begin{equation}
\begin{small}
\begin{array}{l}
D_{j} = \left[ {\begin{array}{*{20}{c}}
{x_1^{{\rm{rank}}1}}&{x_2^{{\rm{rank}}1}}& \cdots &{x_{\|I_j\|}^{{\rm{rank}}1}}&{{{\hat y}^{{\rm{rank}}1}}}\\
{x_1^{{\rm{rank}}2}}&{x_2^{{\rm{rank2}}}}& \cdots &{x_{\|I_j\|}^{{\rm{rank2}}}}&{{{\hat y}^{{\rm{rank2}}}}}\\
 \vdots & \vdots &{}& \vdots & \vdots \\
{x_1^{{\rm{rank}}M}}&{x_2^{{\rm{rank}}M}}& \cdots &{x_{\|I_j\|}^{{\rm{rank}}M}}&{{{\hat y}^{{\rm{rank}}M}}}
\end{array}} \right]
\end{array}
\end{small}
\end{equation}
where, $\|I_j\|$ is the dimension of $j^{th}$ design space, $\hat y$ is the predicted function value using HK, and the superscript $rankM$ indexes the selected sample.

(3) Let the selected local candidate set contain $M$ points from the $N$ sampled points, with $M\leq N$. Thereby, the lower and upper bound of the new searching space can be determined as follows:
\begin{equation}
\begin{small}
\begin{aligned}
    {l_k} = \min {\left[ {x_k^{{\rm{rank}}1},\;\;\;{\mkern 1mu} x_k^{{\rm{rank}}2},\;\;\;{\mkern 1mu}  \cdots ,\;\;\;{\mkern 1mu} x_k^{{\rm{rank}}M}} \right]^{\rm{T}}}\\
{u_k} = \max {\left[ {x_k^{{\rm{rank}}1},\;\;\;{\mkern 1mu} x_k^{{\rm{rank}}2},\;\;\;{\mkern 1mu}  \cdots ,\;\;\;{\mkern 1mu} x_k^{{\rm{rank}}M}} \right]^{\rm{T}}}
\end{aligned}
\end{small}
\end{equation}
where, $l_k$ and $u_k$ are the lower and upper bound value of the $k^{th}$ variable in the $j^{th}$ subset design space. The new searching space of the $j^{th}$ subset design space can be formulated as:
\begin{equation}
\begin{small}
{S_j} = \left\{ {\begin{array}{*{20}{c}}
{{[0,1]}^{\|{I_j}\|}} \qquad \qquad \qquad \qquad \qquad \quad  ({\rm{if}}\;  EI_{\max } > \Delta)\\
\qquad \qquad \qquad \qquad\\
{{[l_{j},u_{j}]}^{\|{I_j}\|}}
\qquad \qquad \qquad \qquad  \qquad ({\rm{if}}\;  EI_{\max } \leq \Delta)
\end{array}} \right.
\end{small}
\end{equation}

The displayed range-switching definition follows Eq.~(15) of the published article. Numeric candidate-pool sizes and the earlier pseudocode are omitted here because the available descriptions do not give an internally consistent specification of those details. This abridgment does not claim a new implementation or an independent reproduction of the published experiments.

\section{Experimental Study}\label{sec4}

In this section, we test KT-EGO and compare it with the state of the art algorithms on numerical benchmark problems and an engineering problem.
\subsection*{4.1 Experimental settings}
\par 
To show the effectiveness of the proposed algorithm, it is compared against the following three kinds of algorithms:
\par
(1)
The first kind is the model-free algorithms including the classical genetic algorithm (GA)\citep{melanie_introduction_1998}, differential evolution algorithm (DE)\citep{storn_differential_1996} and gaining-sharing knowledge based algorithm (GSK) \citep{mohamed_gaining-sharing_2020},
where the GSK is the most recent human-based algorithm.
For these 3 algorithms, their population size is set to be 50 and the remaining variables are in consistent with the default values shown in the related papers. 
\par
(2)
The second kind is the surrogate assisted genetic algorithms including the incremental kriging-assisted evolutionary algorithm (IKAEA)\citep{zhan_fast_2021}
and generalized surrogate-assisted genetic algorithm (GSGA)~\citep{cai_efficient_2020}.
Note that the IKAEA and GSGA are recently proposed SAEAs for high-dimensional optimization problems.
We follow the default settings in the papers of IKAEA and GSGA for the following tests.
\par
(3)
The third kind is the surrogate-based algorithm, Nash-EGO, which has mentioned in Section 2.1.
Originally, the Nash-EGO starts the optimization search from an random sample point, which is selected as the elite point in the first optimization cycle. 
However, for fair comparison with other compared algorithms, here the elite point of Nash-EGO in the first optimization cycle is selected as the best solution of the initial training samples for KT-EGO, GSGA and other compared algorithms.
In the meantime, the number of subset spaces (also called players in the Nash-EGO) are set to be exactly the same as that of KT-EGO.

\par 
In this paper, the tests for all the benchmark problems are repeated 20 times.
In the $i^{\text{th}}$ test of each benchmark function $(i=1 \cdots 20)$, the initial sample distribution (or called initial population) of all algorithms is the same.
The termination condition is the number of function evaluations (NFE) reaches to 3000;
For the KT-EGO algorithm, the value of parameter $t_{\max}$, the total number of iterations in each subset optimization, is set as 6 times of the variable dimension in each subset design space.
\subsection* {4.2 Benchmark functions}
\par
All these benchmark functions used in this paper are selected from the CEC's 2010 special session~\citep{tang_benchmark_2010}. 
The characteristics of these benchmark functions are summarized in Table~\ref{tab:addlabel-1}.
Detailed benchmark definitions are provided with the published article and its supplemental material (DOI: 10.1080/0305215X.2022.2139374).
\begin{table}[htbp]
  \centering
  \tbl{Benchmark functions}
    {\begin{tabular}{llll}
    \toprule
   Index & Dim & Name & Range\\
    \midrule
    F1    & 30    & {Shifted Elliptic} &$[-10,10]^{30}$  \\
    F2    & 30    & {Shifted Rastrigin}        & $[-5,5]^{30}$ \\
    F3    & 30    & {Shifted Ackley}        &$[-32,32]^{30}$ \\
    \midrule
    F4    & 30    &\tabincell{l}{3-group 5-rotated Elliptic}        &$[-10,10]^{30}$  \\
    F5    & 30    &\tabincell{l}{3-group 5-rotated Rastrigin}       & $[-5,5]^{30}$  \\
    F6    & 30    &\tabincell{l}{3-group 5-rotated Ackley}      &$[-32,32]^{30}$  \\
    \midrule
    F7    & 60    & {Shifted Elliptic }      &$[-10,10]^{60}$  \\
    F8    & 60    & {Shifted Rastrigin}       & $[-5,5]^{60}$  \\
    F9    & 60    & {Shifted Ackley}       &$[-32,32]^{60}$  \\
    \midrule
    F10   & 60    &\tabincell{l}{6-group 5-rotated Elliptic }       & $[-10,10]^{60}$ \\
    F11   & 60    &\tabincell{l}{6-group 5-rotated Rastrigin}      & $[-5,5]^{60}$  \\
    F12   & 60    &\tabincell{l}{6-group 5-rotated Ackley}      &$[-32,32]^{60}$  \\
    \bottomrule
    \end{tabular}}%
  \label{tab:addlabel-1}%
\end{table}%
\par
Note that the variable interactions has a great impact on the performance of the optimization algorithm. 
Therefore, 6 separable and 6 non-separable function are selected for our test.
Specifically, the separable functions are the functions whose design variables are independent.
In contrast, the design variables of the non-separable functions are highly interacted.
\subsection*{4.3 Result of benchmark functions}
\subsubsection*{4.3.1 Comparison result of 30-D benchmark functions}
\par
\begin{figure}[htp]
\centering

\subfigure[F1]{
\begin{minipage}[t]{0.33\linewidth}
\centering
\includegraphics[scale=0.33]{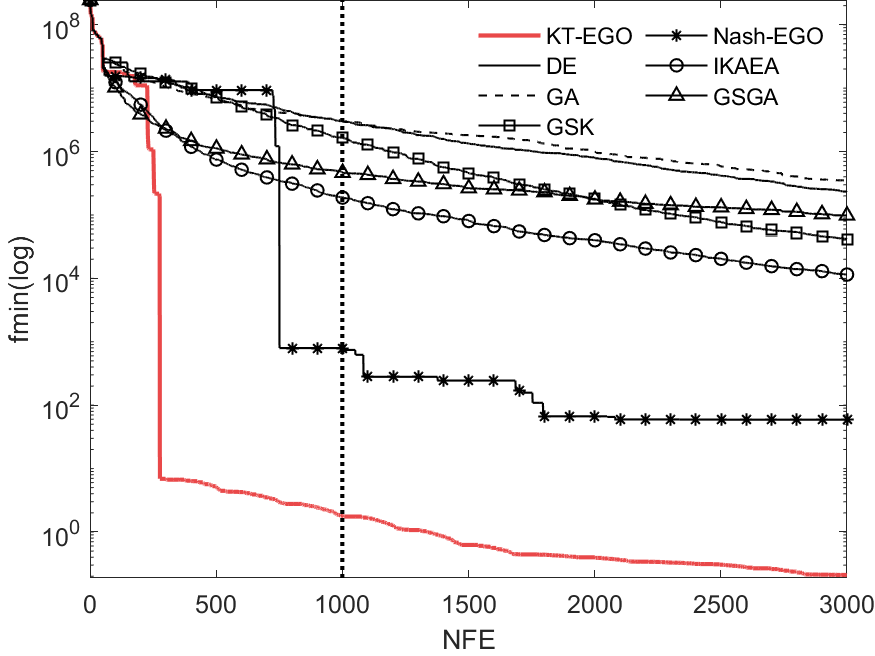}
\end{minipage}%
}%
\subfigure[F2]{
\begin{minipage}[t]{0.33\linewidth}
\centering
\includegraphics[scale=0.33]{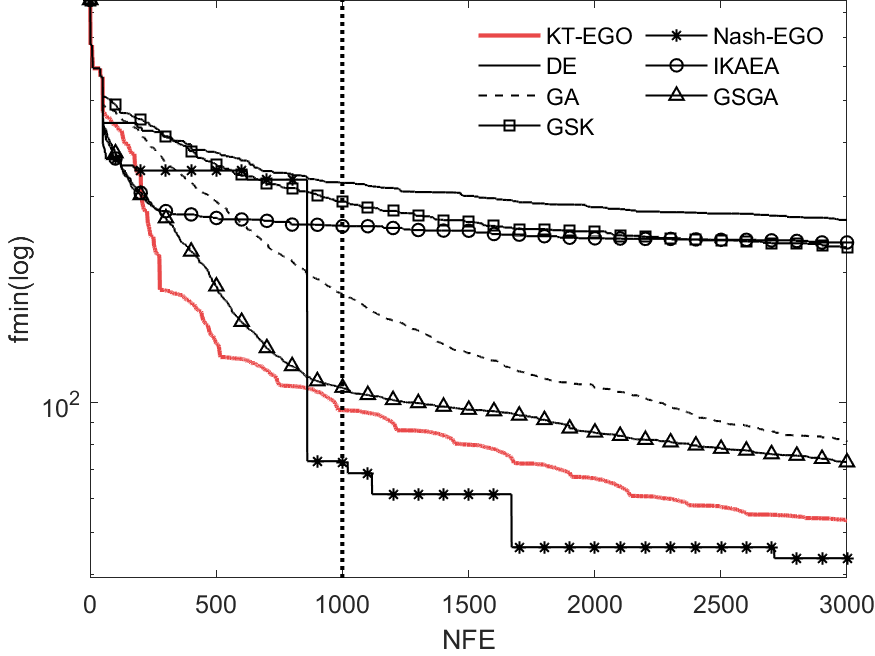}
\end{minipage}
}%
\subfigure[F3]{
\begin{minipage}[t]{0.33\linewidth}
\centering
\includegraphics[scale=0.33]{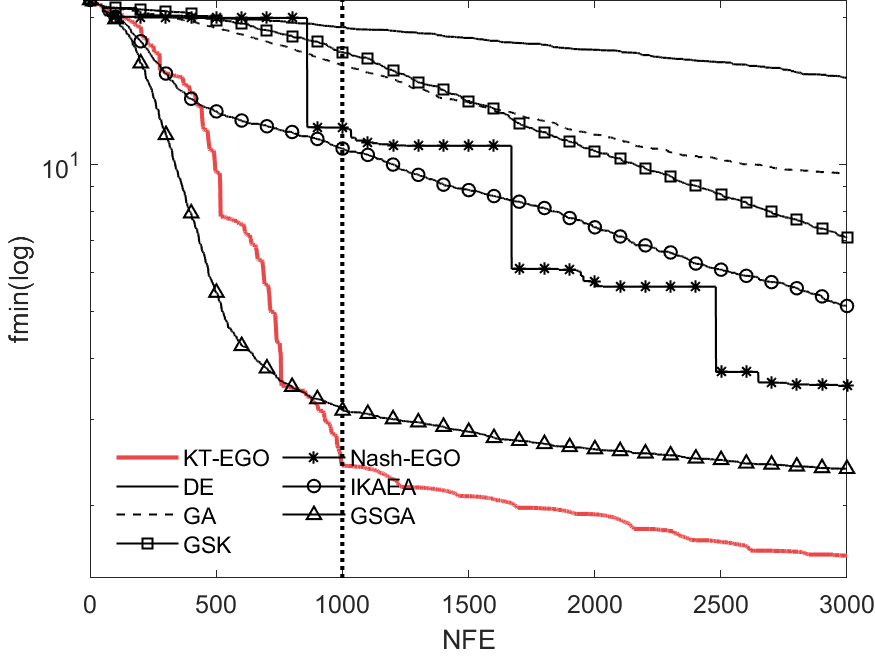}
\end{minipage}
}%

\subfigure[F4]{
\begin{minipage}[t]{0.33\linewidth}
\centering
\includegraphics[scale=0.33]{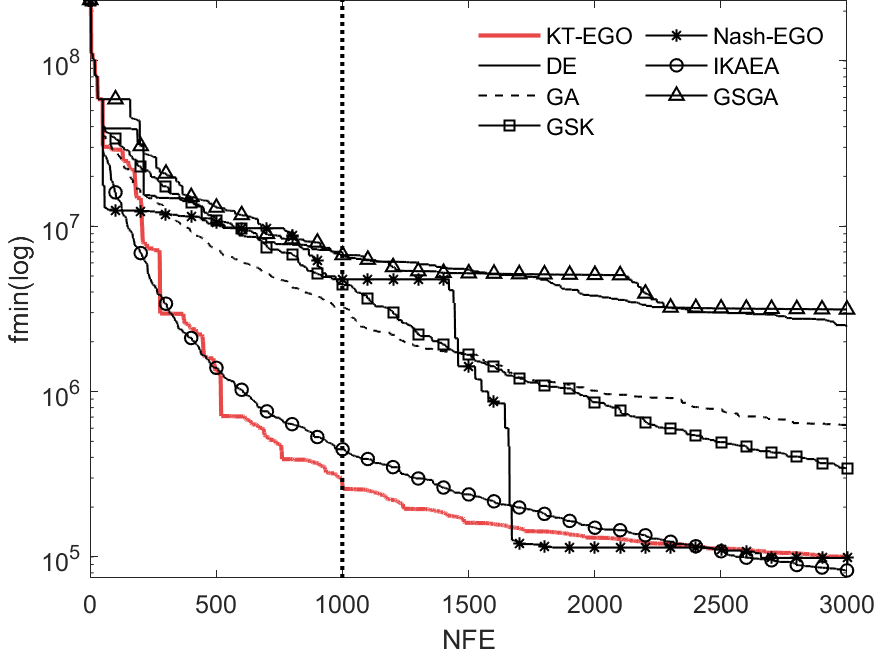}
\end{minipage}%
}%
\subfigure[F5]{
\begin{minipage}[t]{0.33\linewidth}
\centering
\includegraphics[scale=0.33]{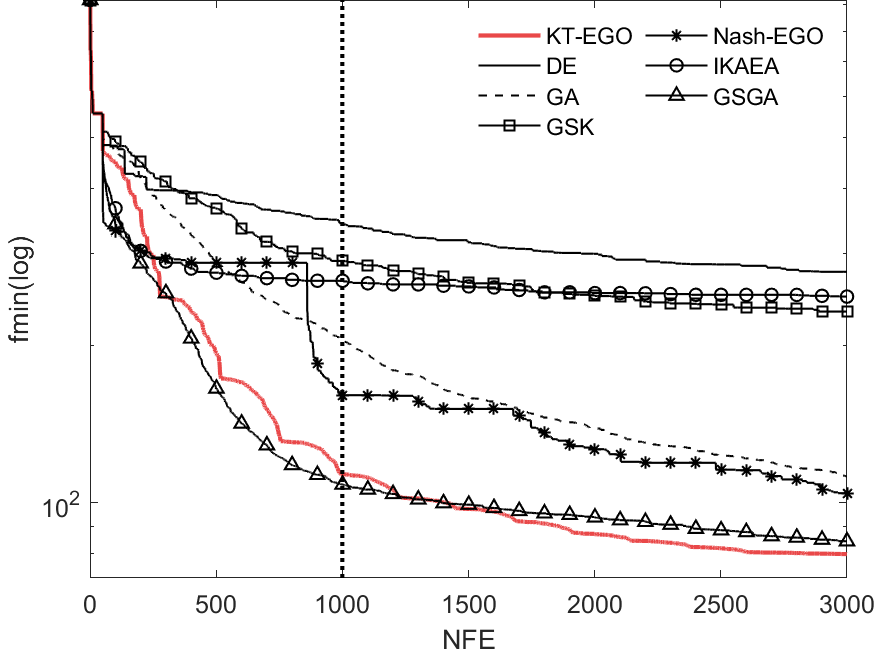}
\end{minipage}
}%
\subfigure[F6]{
\begin{minipage}[t]{0.33\linewidth}
\centering
\includegraphics[scale=0.33]{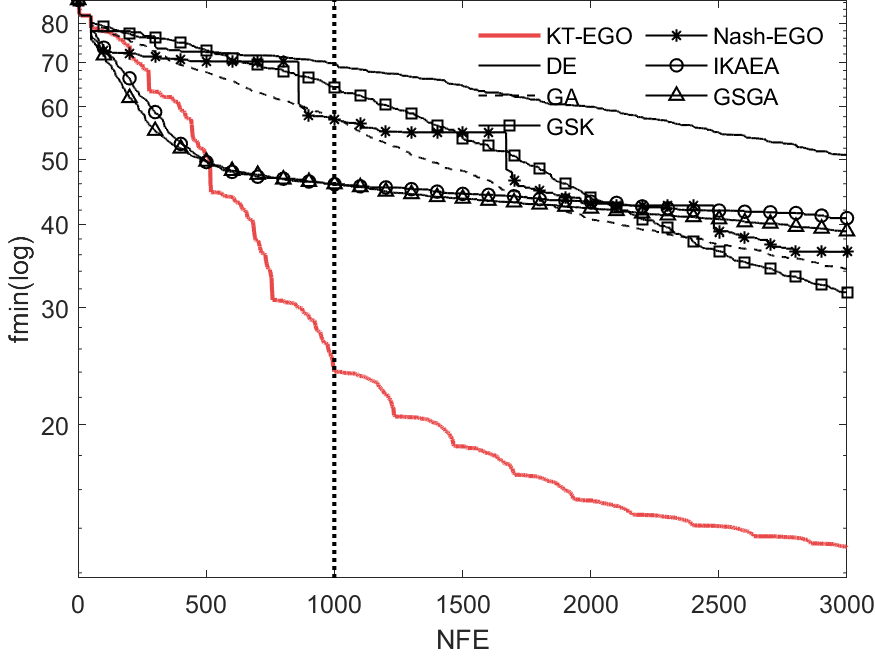}
\end{minipage}
}%

\centering
\caption{The average convergence curves of 7 algorithms on 30-D benchmark functions}
\label{fig:kt-original-5}
\end{figure}

 
The convergence curves of 6 different 30-dimensional benchmark functions are given in Figure~\ref{fig:kt-original-5}.
There is a dashed line indicating 1000 NFE in each figure.
Considering the difference in the number of test samples in different literature, but also in order to present a comprehensive comparison, 
the detailed optimization results at 1000 and 3000 NFE are provided in the published supplemental data,
where ``Std'' represents standard deviation, and the ``Wilcoxon'' represent Wilcoxon signed rank test \citep{derrac_practical_2011}, the symbols `+', `$-$', and `$\approx$' indicate the KT-EGO algorithm is significantly better than, significantly worse than , or comparable to the compared algorithm.
and the optimal mean result of each function is marked in bold.
\par
The statistical results show that the KT-EGO algorithm performs significantly better than the other compared algorithms in most test cases.
As shown in Figure~\ref{fig:kt-original-5}(a), (b), and (c), F1, F2, and F3 are separable functions, 
the decomposition based algorithms such as Nash-EGO and KT-EGO have advantages over the other compared baselines. Because in these benchmark functions, the combination of all the sub-optimization results is bound to get a better solution in the global space.
In the meantime, F1 is also a uni-modal function, which can be accurately fitted by the PCE surrogate with very limited samples. 
So, the KT-EGO algorithm shows a great advantage in the early stage.
In multi-modal functions F2 and F3, the GSGA algorithm also has great performance.
\par 
On the other hand, F4, F5, and F6 are non-separable functions.
For such problems, the performance of Nash-EGO becomes worse when compared to the testing results on separable functions. In contrast, the convergence rate and the final solutions of KT-EGO are better than the compared algorithms, especially when optimizing the F6 functions.
The values of standard deviation reflect the robustness of the algorithm for the initial distribution.%

\subsubsection*{4.3.2 Comparison results of 60-D benchmark functions}
\par
\begin{figure*}[htbp]
\centering

\subfigure[F7]{
\begin{minipage}[t]{0.33\linewidth}
\centering
\includegraphics[scale=0.33]{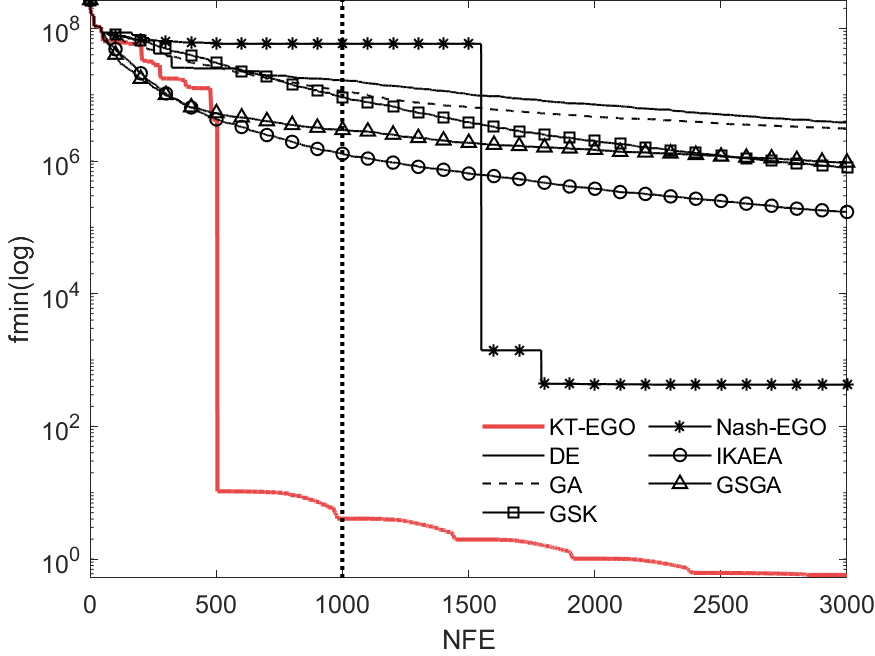}
\end{minipage}%
}%
\subfigure[F8]{
\begin{minipage}[t]{0.33\linewidth}
\centering
\includegraphics[scale=0.33]{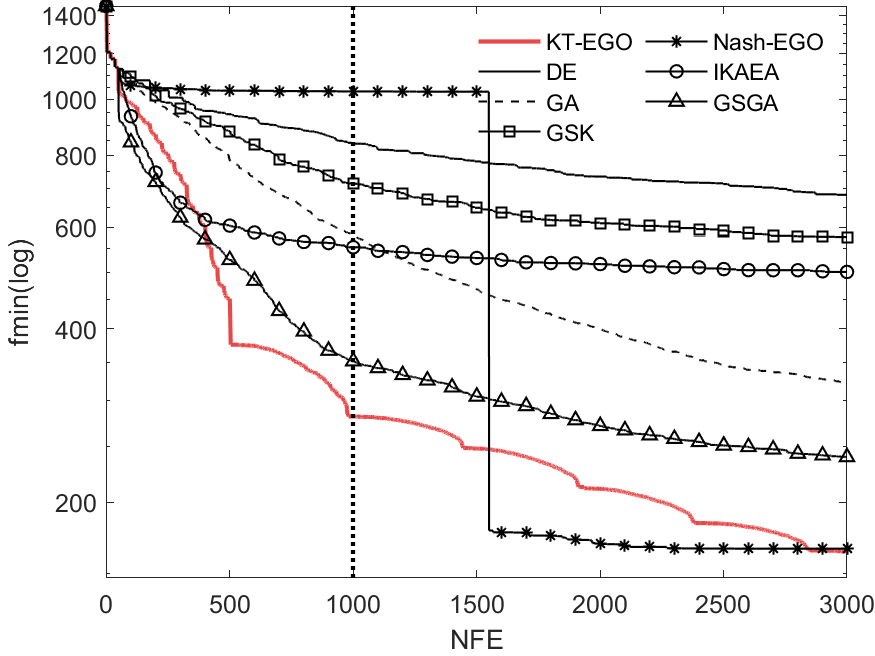}
\end{minipage}
}%
\subfigure[F9]{
\begin{minipage}[t]{0.33\linewidth}
\centering
\includegraphics[scale=0.33]{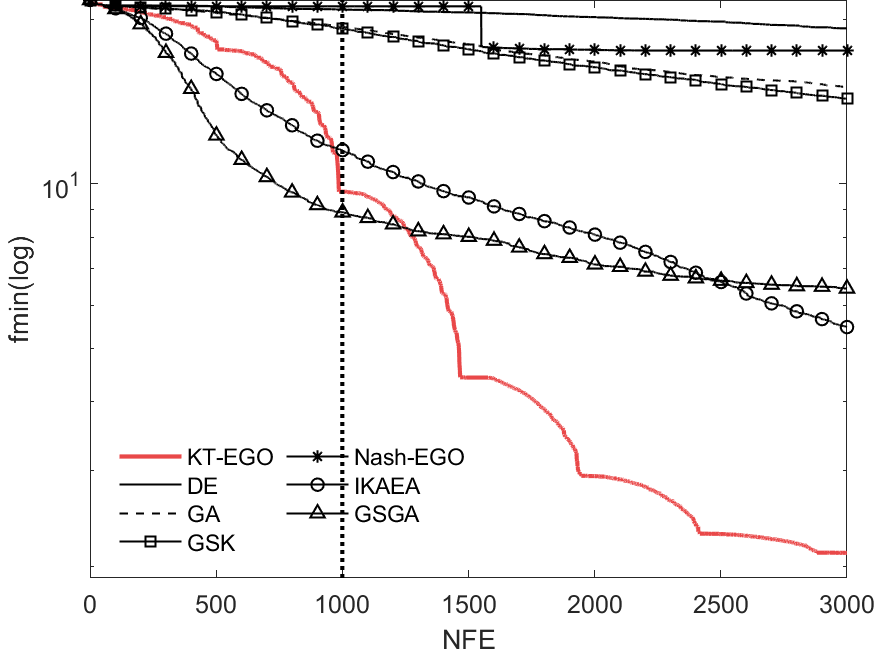}
\end{minipage}
}%

\subfigure[F10]{
\begin{minipage}[t]{0.33\linewidth}
\centering
\includegraphics[scale=0.33]{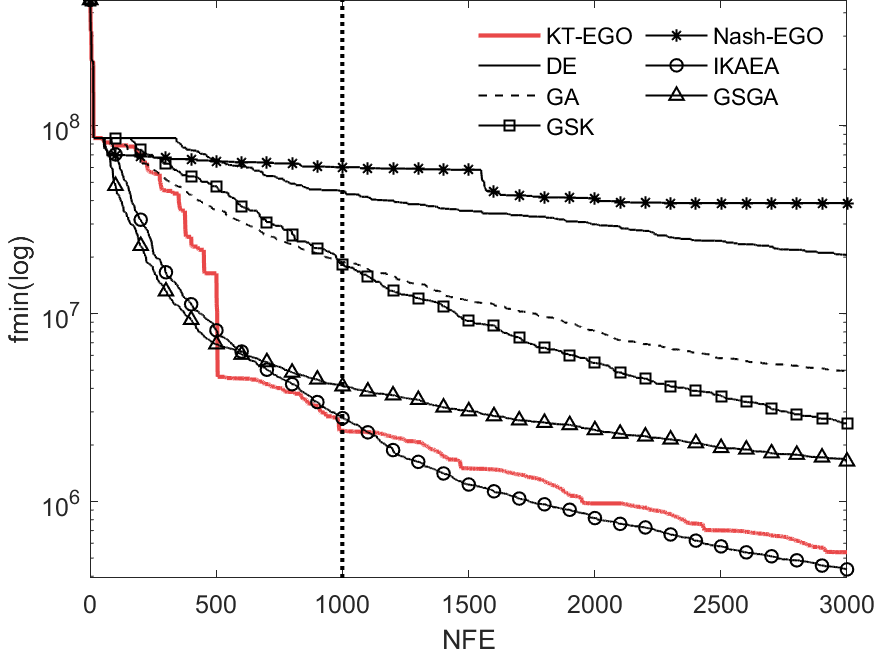}
\end{minipage}%
}%
\subfigure[F11]{
\begin{minipage}[t]{0.33\linewidth}
\centering
\includegraphics[scale=0.33]{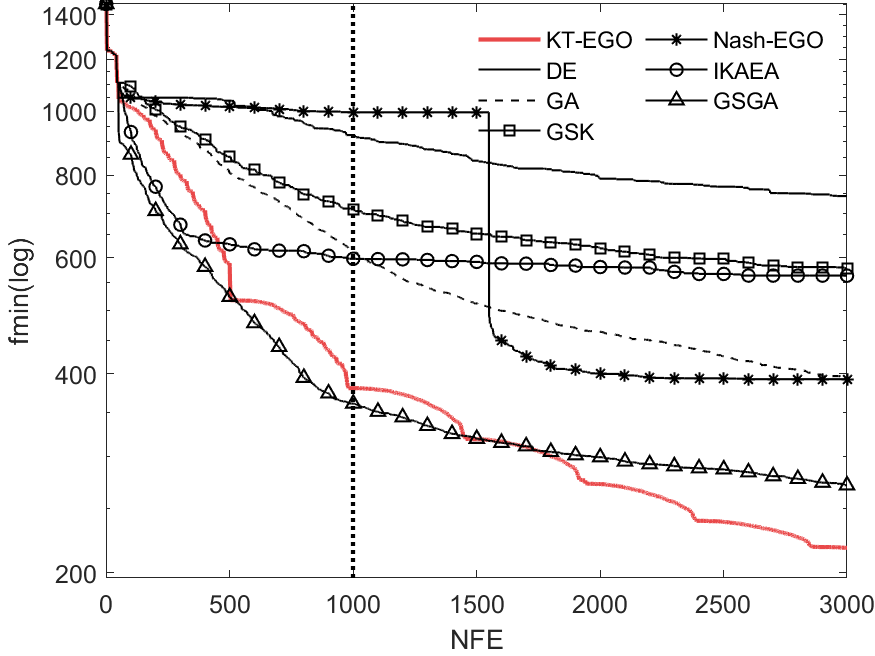}
\end{minipage}
}%
\subfigure[F12]{
\begin{minipage}[t]{0.33\linewidth}
\centering
\includegraphics[scale=0.33]{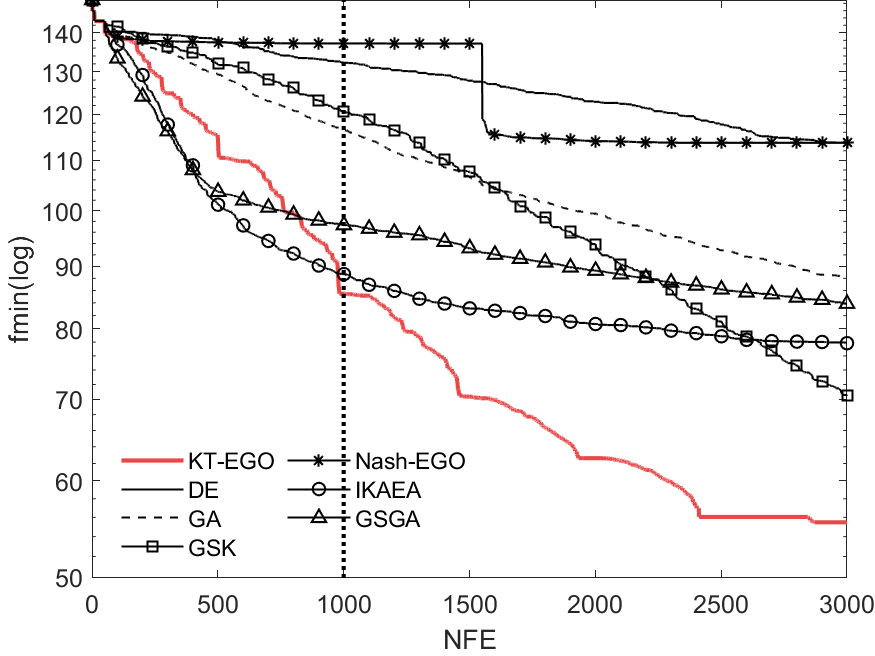}
\end{minipage}
}%
\centering
\caption{The average convergence curves of 7 algorithms on 60-D benchmark functions}
\label{fig:kt-original-6}
\end{figure*}

The averaged convergence curves of 7 algorithms on six 60-dimensional benchmark functions are given in Figure~\ref{fig:kt-original-6}.
Detailed results are provided in the published supplemental data.
The statistical results show that the KT-EGO algorithm performs significantly better than or is comparable with other compared algorithms in all test cases.

\par 
Similar to the 30-D benchmark function tests' result, 
the KT-EGO algorithm has an obvious advantage on separable benchmark functions F7$\sim$F9.
When optimizing the non-separable functions such as F10, F11, and F12, the performance of KT-EGO is still competitive.
Though the convergence rates of IKAEA can be better than those of KT-EGO at the early stage of the optimization process, the KT-EGO usually achieves better final solutions, as shown in Figure~\ref{fig:kt-original-6}(e) and (f).
The final mean results support the effectiveness of KT-EGO on these benchmark cases.
\subsection*{4.4 Discussion of proposed algorithm }
After the effectiveness of the KT-EGO has been verified with tests on 12 benchmark functions, the proposed algorithm will be discussed in more detail in this section.
First, two variants algorithms of KT-EGO are tested to demonstrate the effectiveness of the two proposed strategies. 

\subsubsection*{4.4.1 Effectiveness of the KT-EGO algorithm components}
\par
\begin{figure*}[htbp]
\centering
\subfigure[F1]{
\begin{minipage}[t]{0.33\linewidth}
\centering
\includegraphics[scale=0.33]{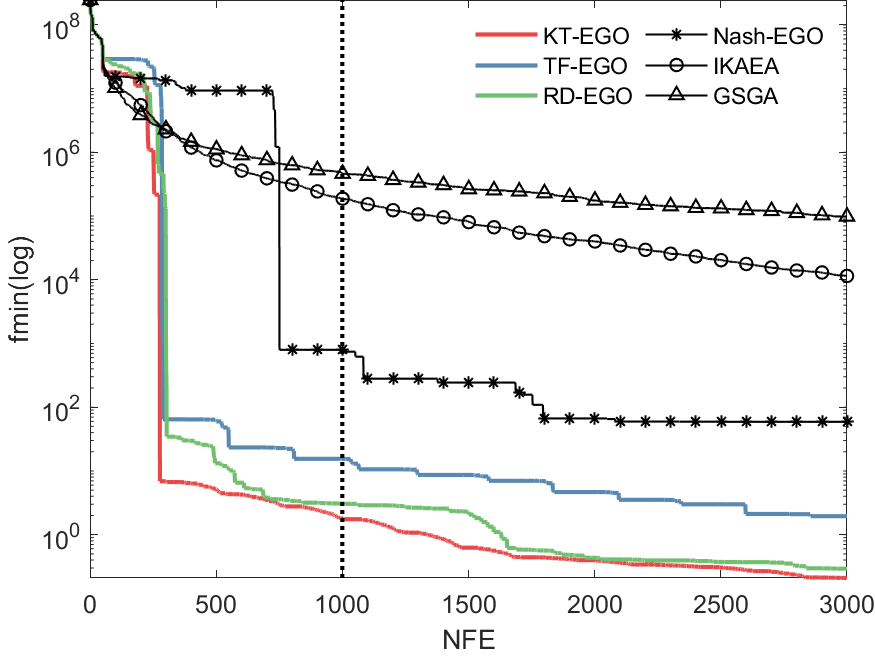}
\end{minipage}%
}%
\subfigure[F2]{
\begin{minipage}[t]{0.33\linewidth}
\centering
\includegraphics[scale=0.33]{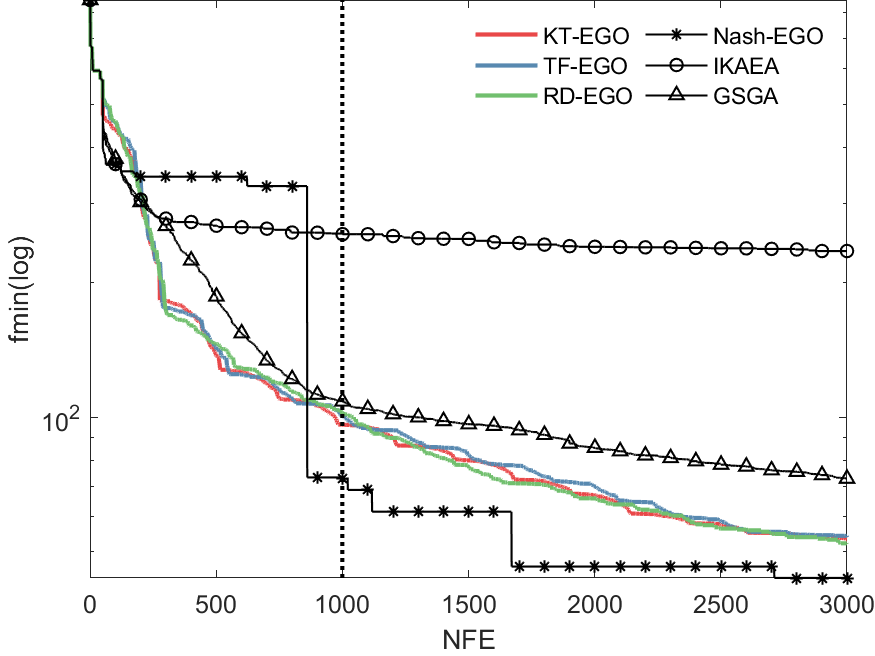}
\end{minipage}
}%
\subfigure[F3]{
\begin{minipage}[t]{0.33\linewidth}
\centering
\includegraphics[scale=0.33]{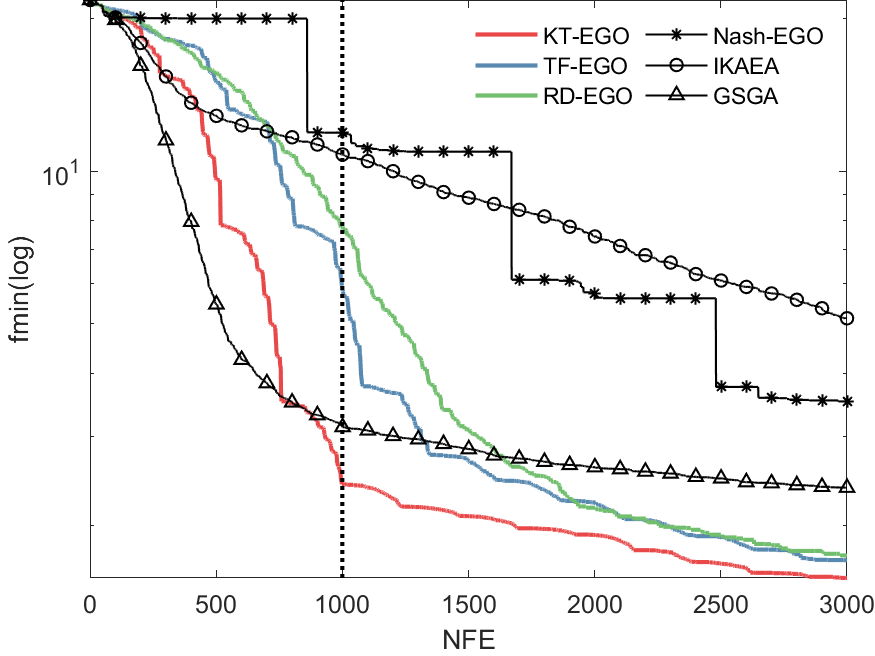}
\end{minipage}
}%

\subfigure[F4]{
\begin{minipage}[t]{0.33\linewidth}
\centering
\includegraphics[scale=0.33]{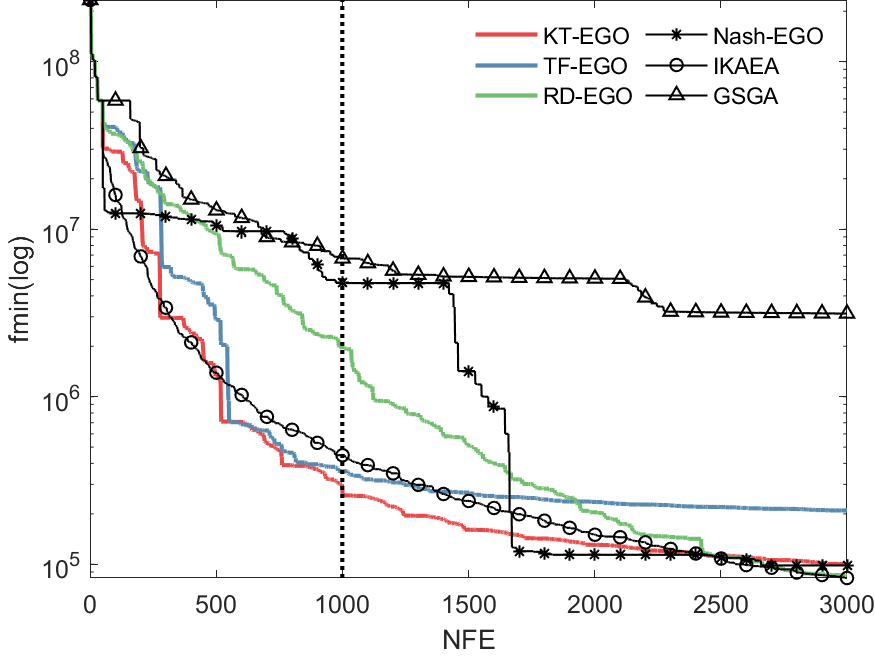}
\end{minipage}%
}%
\subfigure[F5]{
\begin{minipage}[t]{0.33\linewidth}
\centering
\includegraphics[scale=0.33]{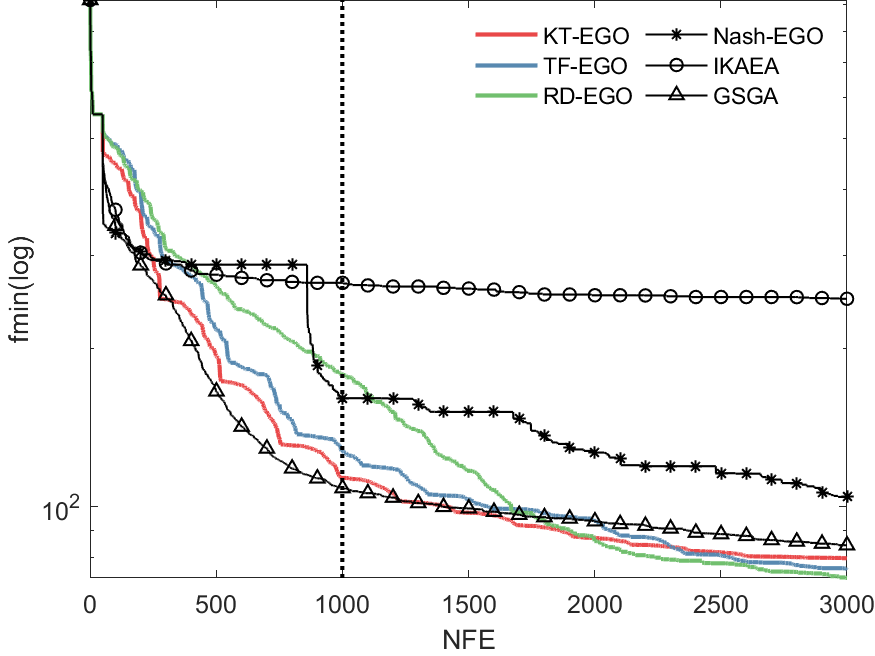}
\end{minipage}
}%
\subfigure[F6]{
\begin{minipage}[t]{0.33\linewidth}
\centering
\includegraphics[scale=0.33]{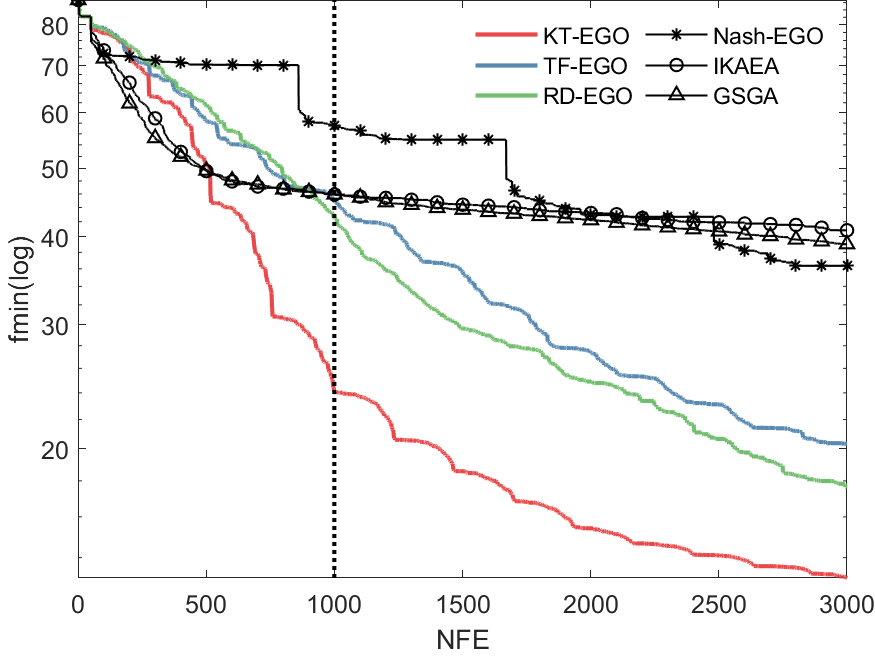}
\end{minipage}
}%
\centering
\caption{The average convergence curves of the KT-EGO algorithm and its 2 variants algorithms.}
\label{fig:kt-original-7}
\end{figure*}

 
Section 3.1 and 3.2 introduced the components of the KT-EGO algorithm, which are the random decomposition strategy and the knowledge transfer strategy.
To verify the effectiveness of these 2 components,
2 variant algorithms of the KT-EGO, RD-EGO and TF-EGO, are constructed,
which corresponds to the above 2 strategies respectively.
Detailed results of these algorithms are provided in the published supplemental data.
Figure~\ref{fig:kt-original-7} shows the average convergence curves of the KT-EGO algorithm and its 2 variants algorithms, and the convergence curves of the other 3 algorithms are also drawn as references.
\vspace{0.5cm}

For the random decomposition strategy, the performance of RD-EGO is significantly better than other compared algorithms in most benchmark functions. However, the effectiveness of the decomposition strategy is not so obvious in the 1000 NFE condition.  
This shows that the random decomposition strategy alone can not accelerate the convergence at the initial stage of optimization.
From another point of view, the KT-EGO is more efficient than the TF-EGO, which means that the random grouping strategy indeed improves the accuracy of the global PCE surrogate through more uniform distribution of samples.
\par
For the knowledge transfer strategy, the performance of TF-EGO is also significantly better than other compared algorithms.
The performance of the TF-EGO is similar to KT-EGO in the former stage but opened the gap at the later stage.
The reason for the above phenomenon is that the effectiveness of knowledge transfer mainly accelerates global exploration.
And it is more difficult for interactive variables to be grouped in the same group without the random grouping strategy. 
\par 
The test results of two variant algorithms demonstrate the effectiveness of both two components of the proposed KT-EGO algorithm.

\subsection*{4.5 Engineering test case}
\par 
In this section, the proposed KT-EGO optimization algorithm is applied to the aerodynamic optimization of Rotor37 blades. The optimization results of rotor37 are compared with four other algorithms. 
Each algorithm is optimized 5 times independently, and the initial distribution of 50 samples for each optimization case is generated by the Matlab built-in function "lhsdesign".
\subsubsection*{4.5.1 Problem description}
The compressor is a turbomachine that converts mechanical energy into fluid kinetic energy and potential energy. 
As an important part of a gas turbine, the performance of the compressor has a direct impact on the efficiency, power, and reliability of the gas turbine. 
The optimal design of the compressor is difficult, as the shape of the compressor blade is complex and the profile varies greatly with the blade height. We select the well-known Rotor 37 blade~\citep{suder_experimental_1996} as the design object.
\par
\begin{figure}[ht]
\begin{center}
\includegraphics[width=\linewidth, trim = 0 0.5cm 0 0.5cm]{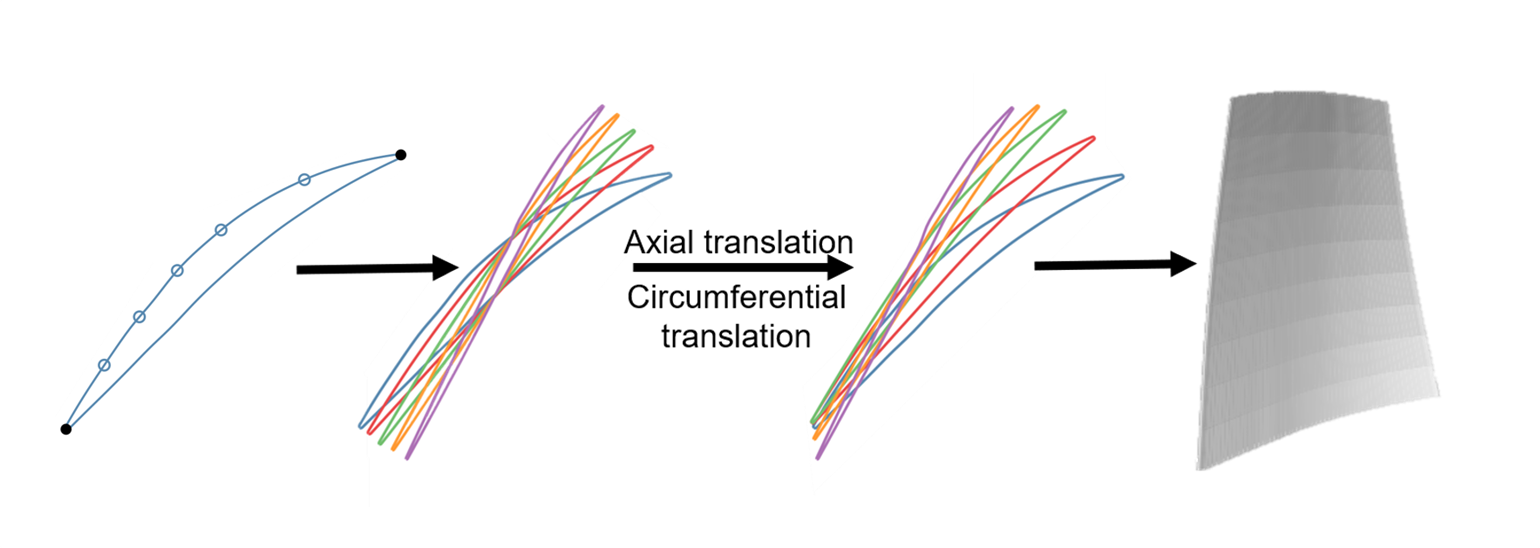}
\end{center}
\caption{ The sketch map of the 3D parameterization method }   
\label{fig:kt-original-8}
\end{figure}

Due to the strong three-dimensional effect of the flow in the compressor, the shape of the compressor blade is usually very complex, and the profile shape from the root to the tip varies greatly, with obvious bending and sweeping at the same time.
Therefore, it is necessary to select more sections to accurately control the shape of the compressor blade. 
Here we select 5 section profiles of 0\%, 25\%, 50\%, 75\%, and 100\% span, and  5 active control points at the suction side of each section are selected to adjust the section profiles. 
At the same time, the bending and sweeping of the blade are also adjusted. 
The number of design variables in the optimization of Rotor 37 is 28, these variables are shown in Table~\ref{tab:addlabel-8}.
Figure~\ref{fig:kt-original-8} shows the generation of three-dimensional blade geometric modeling. 
The parameters of the 5 control points of each section determine the shape of this section. 
After finishing the parameterization of all section profiles, 1 circumferential translation parameter $x_{27}$ for the middle section and 2 axial translation parameters $x_{26},x_{28}$ for the tip section are selected to adjust the stacking line in 3D space.
Then, a 3D blade profile is obtained using skinning surface techniques.
\begin{table}[htbp]
  \centering
  \caption{Design variables in engineering test case}
    \begin{tabular}{cc}
    \toprule
   Geometric definition & variable index \\   
        \midrule
    control coefficient in 0\% span & $x_1,\cdots,x_5$ \\
    control coefficient in 25\% span &$x_6,\cdots,x_{10}$  \\
    control coefficient in 50\% span & $x_{11},\cdots,x_{15}$  \\
    control coefficient in 75\% span &$x_{16},\cdots,x_{20}$\\
    control coefficient in 100\% span & $x_{21},\cdots,x_{25}$  \\
    circumferential translation parameter & $x_{27}$  \\
    axial translation parameter & $x_{26},x_{28}$  \\
        \bottomrule
    \end{tabular}%
  \label{tab:addlabel-8}%
\end{table}%
The isentropic efficiency $\eta _{is}$ is set as the objective function for the optimization, the related optimization model is shown below:
\begin{equation}
\begin{array}{c}
\max  \{f_{obj}(\mathbf{x})\}=\max\{\eta_{is}(\mathbf{x})/p(\mathbf{x})\} \\
\text { s.t. }  0.98 \cdot m(\text {ref}) \leq m(\mathbf{x}) \leq 1.02 \cdot m(\text {ref}) \\
\text{where    }{\eta _{is}} = 
{
\{\left( {p_{outlet}^t}/{p_{inlet}^t} \right)
^\frac{\gamma-1}{\gamma} - 1}
\}/
{
\left({{T_{outlet}^t}/{T_{inlet}^t} - 1}\right)
}
\end{array}
\end{equation}
where,  $\text {ref}$ means the reference design, $m$ is the mass flow rate, ${p_{outlet}^t}/{p_{inlet}^t}$ and ${T_{outlet}^t}/{T_{inlet}^t}$ are the total pressure ratio and total temperature ratio respectively, and $\gamma$ denotes adiabatic exponent.
In the meantime, the design point flow of the optimized design is constrained so that its change does not exceed 2\% of the reference design mass flow. The constraint is realized in the form of penalty function $p(x)$ ~\citep{song_research_2016}.

\subsubsection*{4.5.2 Numerical simulation model}
\par
The Rotor37 blade is one of the rotors of the four-stage axial flow compressor Stage 37 with a high-pressure ratio. It was designed and tested by Reid and Moore in the NASA Glenn center in the 1970s. The design parameters are taken as the inlet stage parameters of a typical aero-engine. 
Table~\ref{tab:addlabel-9} shows the relevant design parameters of Rotor37, which keep the same as the literature~\citep{boretti_experimental_2010}. 
To be in accordance with the literature, uniform total pressure and temperature are imposed at the inlet boundary, and an averaged static pressure is imposed at the outlet. The optimization is carried out with a constant outlet static pressure of 115000 Pa, corresponding to a relative mass flow rate of 99\% for the Rotor 37 blade.
\begin{table}[htbp]
  \centering
  \caption{Design conditions}
    \begin{tabular}{cc}
    \toprule
    Condition name & Value \\
    \midrule
    equivalent rotational speed[rpm] & 17188.7 \\
    number of rotor blades & 36 \\
    rotor blade aspect ratio & 1.19 \\
    tip clearance gap[mm] & 0.356 \\
    inlet total temperature[K] & 288.15 \\
    inlet total pressure[Pa] & 101325 \\
    \bottomrule
    \end{tabular}%
  \label{tab:addlabel-9}%
\end{table}%
For the above model, a grid with about  $4 \times {10^{ 6}}$ nodes is used to calculate. 
The H–O–I topology is employed for the generation of the grid by using the commercial software NUMECA auto-grid5, and the mesh thickness of the first layer near the wall is set to be $3 \times {10^{ - 6}}$ m.

In the calculation, the Spalart-Allmaras turbulence model is used with adiabatic smooth walls condition. 
The Reynolds-averaged Navier–Stokes equations are solved by using the commercial software NUMECA FINE/TURBO. 
With the CPU Intel(R) i5-9400F@2.9GHz, the calculation time of a single sample is about 600s. 
\begin{figure}[ht]
\begin{center}
\includegraphics[scale=0.35, trim = 7cm 5cm 11cm 3.5cm,clip]{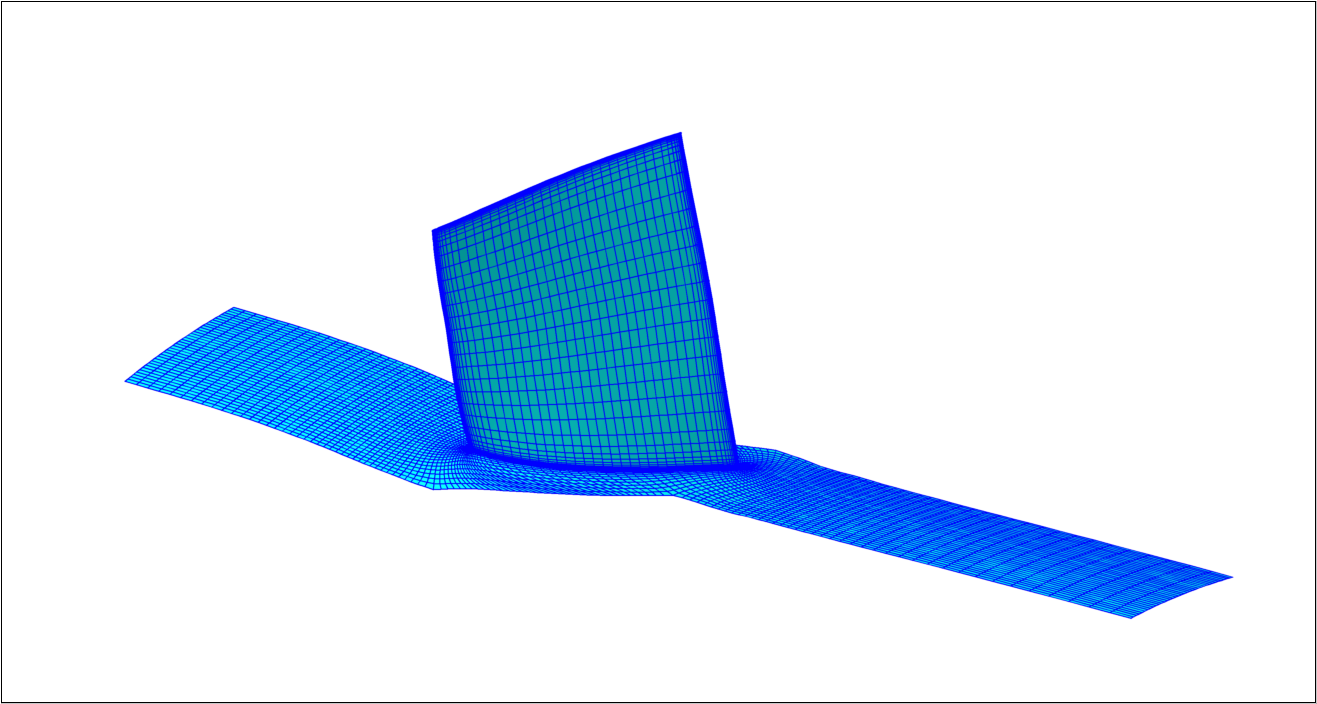}
\end{center}
\caption{The mesh of Rotor 37 blade with $4 \times {10^{ 6}}$ nodes}   
\label{fig:kt-original-9}
\end{figure}
In the design condition, the efficiency, pressure ratio, and mass flow of the reference design are 85.39\%, 2.0395, and 20.62kg/s respectively.

\subsubsection*{4.5.3 Results Analysis}
\par
These algorithms are used to optimize the Rotor 37 blade 5 times, and the total number of CFD calculations in each optimization is 1000.
The published engineering results are reproduced in Table~\ref{tab:addlabel-10}.

\begin{figure}[ht]
\begin{center}
\includegraphics[scale=0.7, trim = 0 0 0 0]{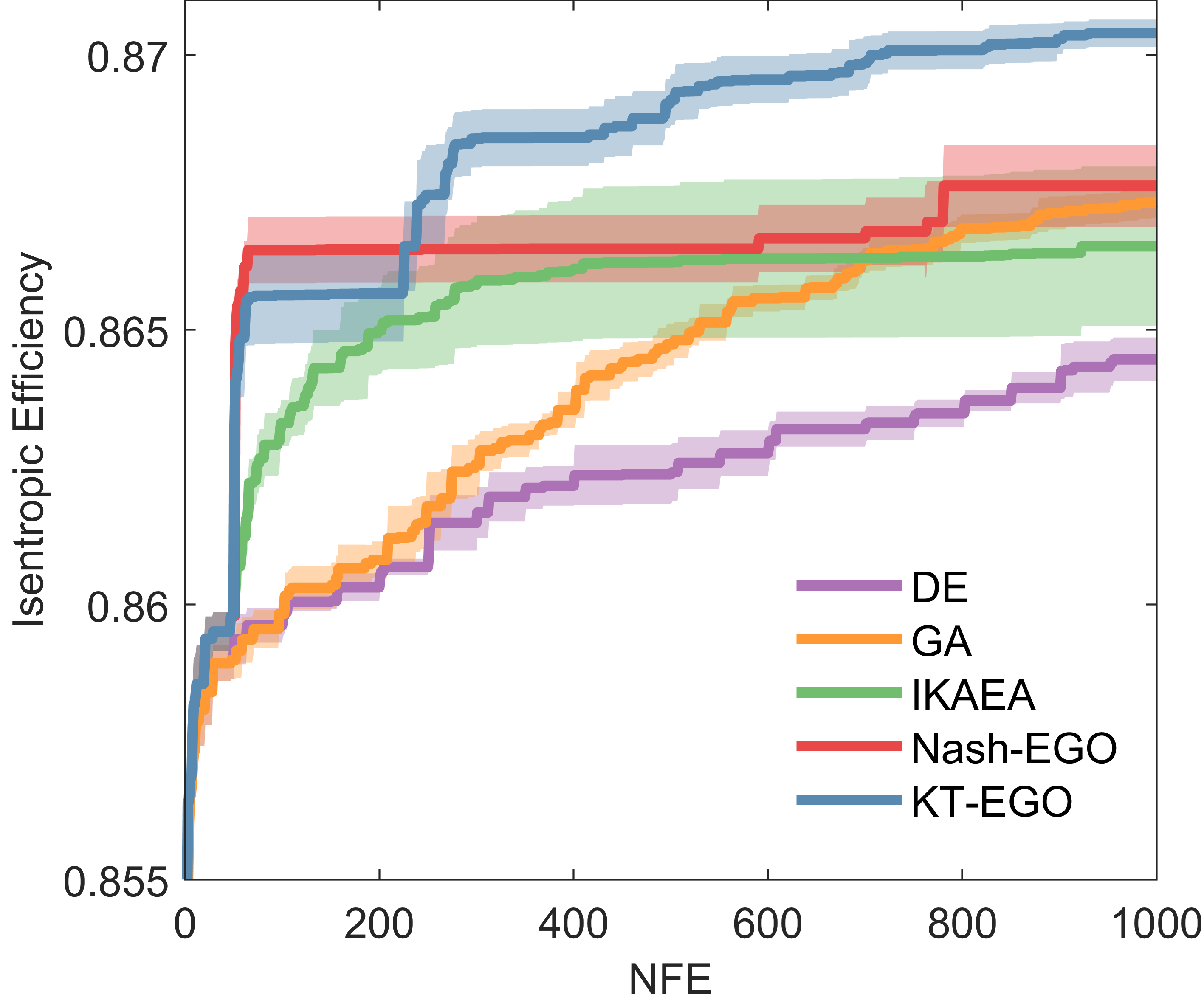}
\end{center}
\caption{The optimization convergence curves of the engineering test case.}   
\label{fig:kt-original-10}
\end{figure}
\begin{table*}[htbp]
  \centering
\tbl{The optimization results of the engineering test case in detail.}
    {\begin{tabular}{lcccc}
\toprule
    Algorithm & Efficiency  &Pressure ratio &Mass flow(kg/s) &\makecell[c]{Efficiency gain\\(percentage points)}\\
    \midrule
    Baseline & 0.8539 & 2.040 & 20.62 & -- \\
    KT-EGO    & 0.8704 $\pm$ 0.0005 & 2.054 $\pm$ 0.0006 & 20.95 $\pm$ 0.0069 & 1.65 \\
    Nash-EGO  & 0.8676 $\pm$ 0.0015 & 2.051 $\pm$ 0.0012 & 20.89 $\pm$ 0.0157 & 1.37 \\
    IKAEA & 0.8665 $\pm$ 0.0029 & 2.050 $\pm$ 0.0045 & 20.88 $\pm$ 0.0539 & 1.26 \\
    GA    & 0.8673 $\pm$ 0.0006 & 2.049 $\pm$  0.0014 & 20.88 $\pm$ 0.0182 & 1.34 \\
    DE    & 0.8645 $\pm$ 0.0008 & 2.048 $\pm$ 0.0059 & 20.83 $\pm$ 0.0350 & 1.06 \\
    \bottomrule
    \end{tabular}}%
  \label{tab:addlabel-10}%
\end{table*}%
\par
All 5 algorithms improve the efficiency of the blade, among which the optimization result of the KT-EGO algorithm is the best, which has great advantages in convergence efficiency and robustness compared with other algorithms. 
After 1000 times of CFD calculations with the KT-EGO, the total efficiency of Rotor 37 blades has been improved by 1.65 percentage points (0.8539 to 0.8704). The variance of the results of 5 repeated operations of KT-EGO, GA, and DE algorithms is very small, indicating that these algorithms are less affected by the initial distribution of samples.
\par
\begin{figure}[htbp]
\centering
\subfigure[efficiency-mass flow]{
\begin{minipage}[t]{0.45\linewidth}
\centering
\includegraphics[scale=0.50]{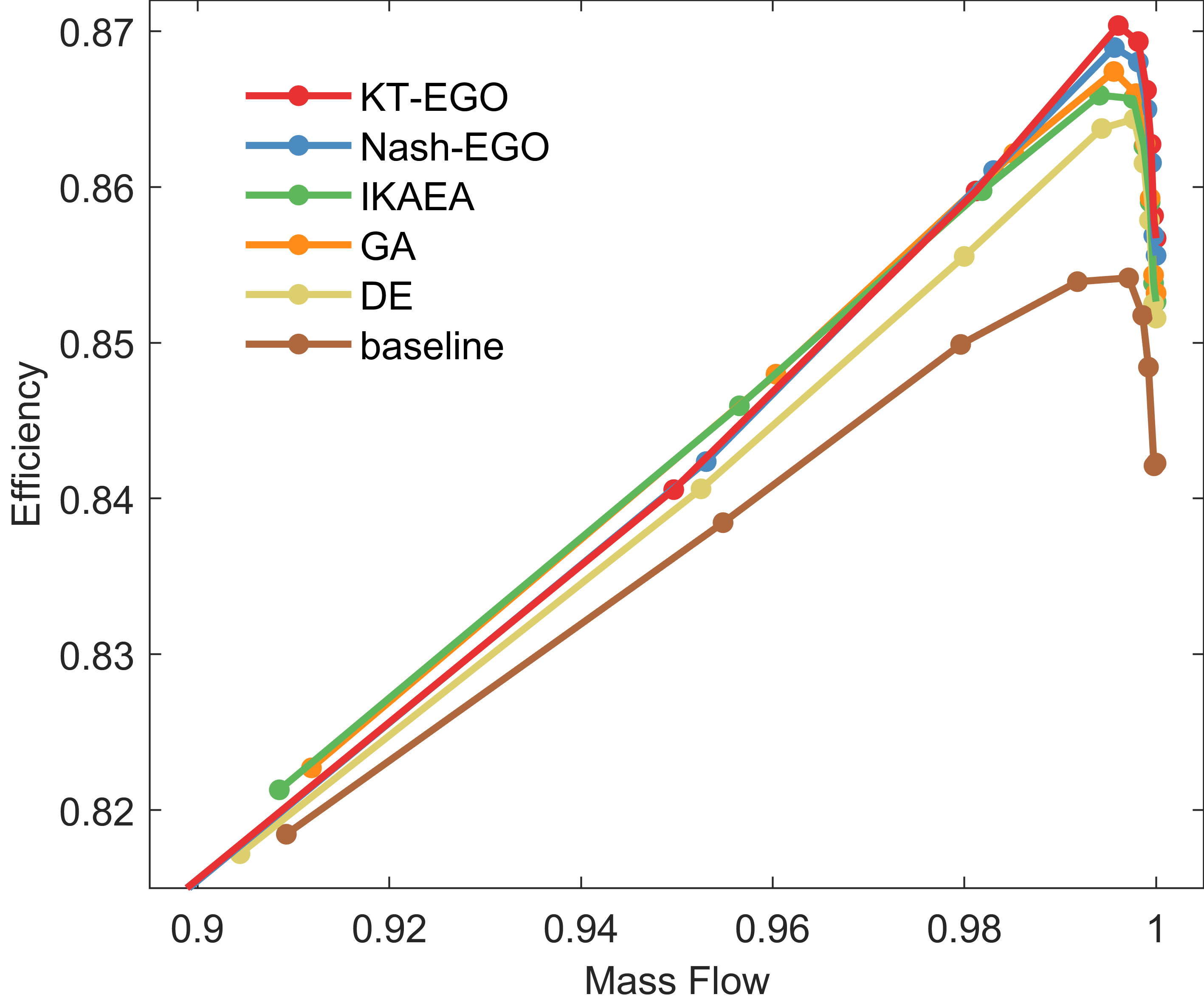}
\end{minipage}%
}%
\subfigure[pressure ratio-mass flow]{
\begin{minipage}[t]{0.45\linewidth}
\centering
\includegraphics[scale=0.50]{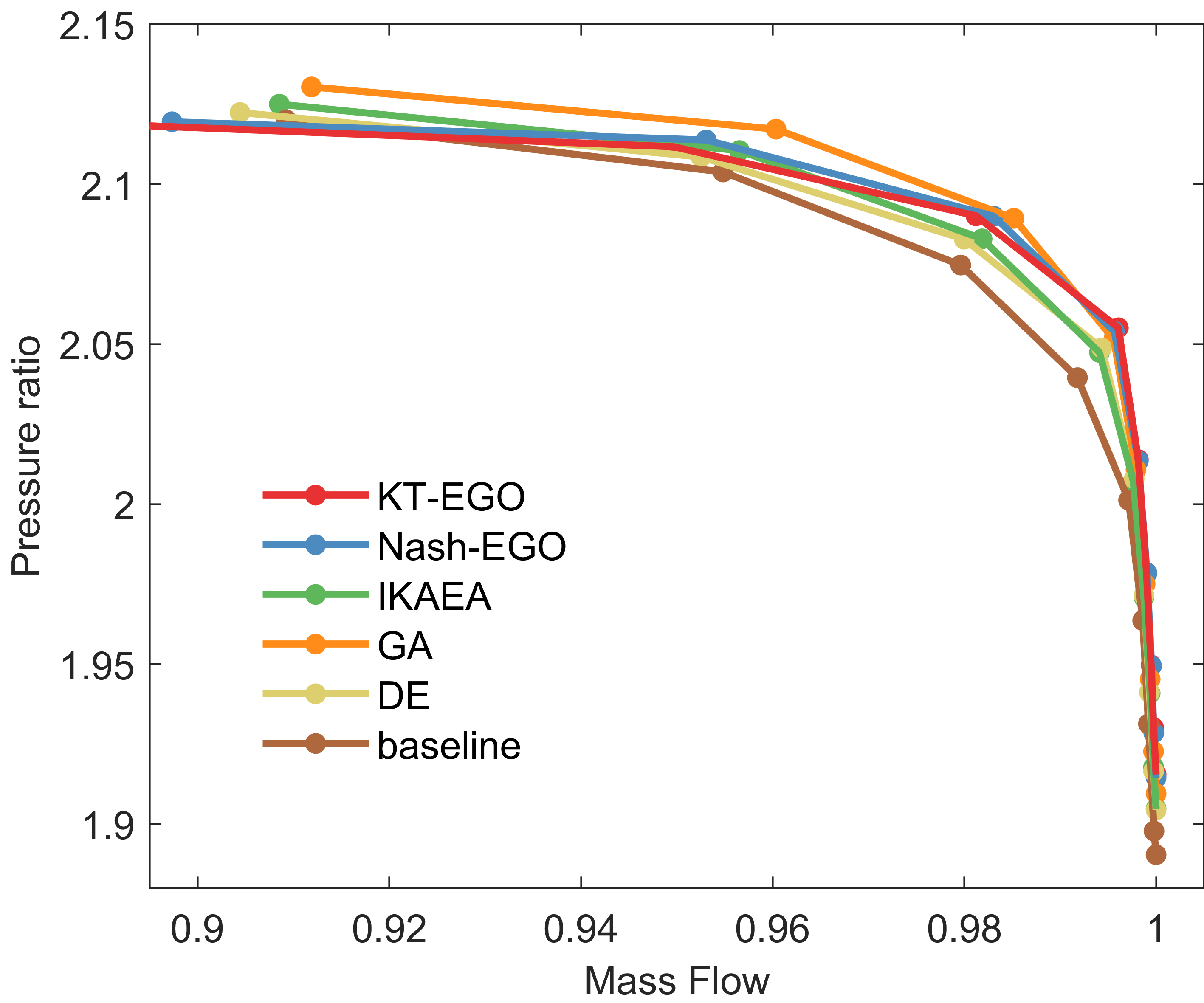}
\end{minipage}
}%
\centering
\caption{The comparison of feature curves of optimization results of different algorithms and the baseline design }
\label{fig:kt-original-11}
\end{figure}

 
Figure~\ref{fig:kt-original-11} shows the overall performance at off-design conditions of the Rotor37 blade and designs optimized by different algorithms. 
Figure~\ref{fig:kt-original-11}(a) depicts the relationship between Isentropic efficiency and mass flow under different operating conditions, we can see that the optimal design result obtained by the KT-EGO algorithm is more efficient than the reference design in the whole operating range.
Figure~\ref{fig:kt-original-11}(b) describes the relationship between pressure ratio and mass flow under different operating conditions. 
When the relative flow is greater than 96\%, the pressure of the KT-EGO algorithm optimization design is slightly higher than that of the reference design. 
The optimization result of the KT-EGO algorithm not only improves the efficiency in the whole operating range but also ensures the stability of the compressor operation.
\par
\begin{figure}[ht]
\centering
\subfigure[baseline]{
\begin{minipage}[t]{0.33\linewidth}
\centering
\includegraphics[scale=0.26,trim=11cm 0.5cm 11cm 0.2cm,clip]{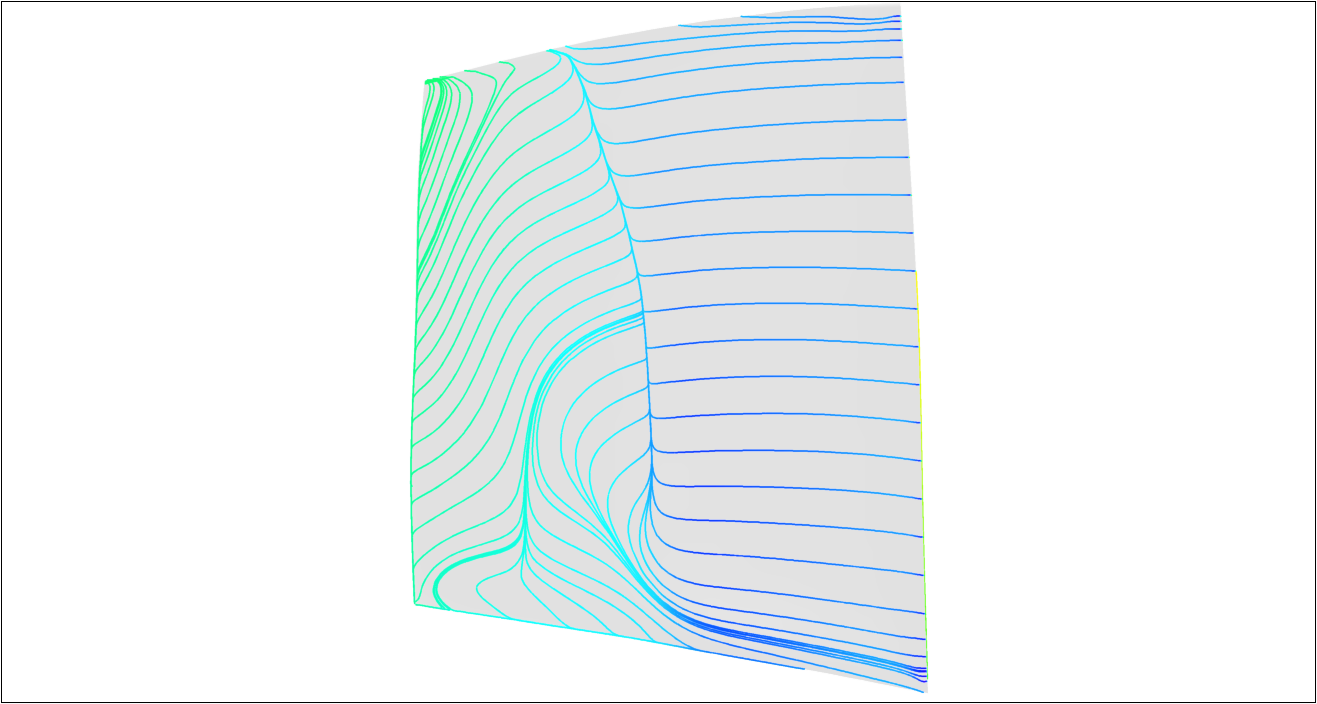}
\end{minipage}%
}%
\subfigure[KT-EGO]{
\begin{minipage}[t]{0.33\linewidth}
\centering
\includegraphics[scale=0.26,trim=11cm 0.5cm 11cm 0.2cm,clip]{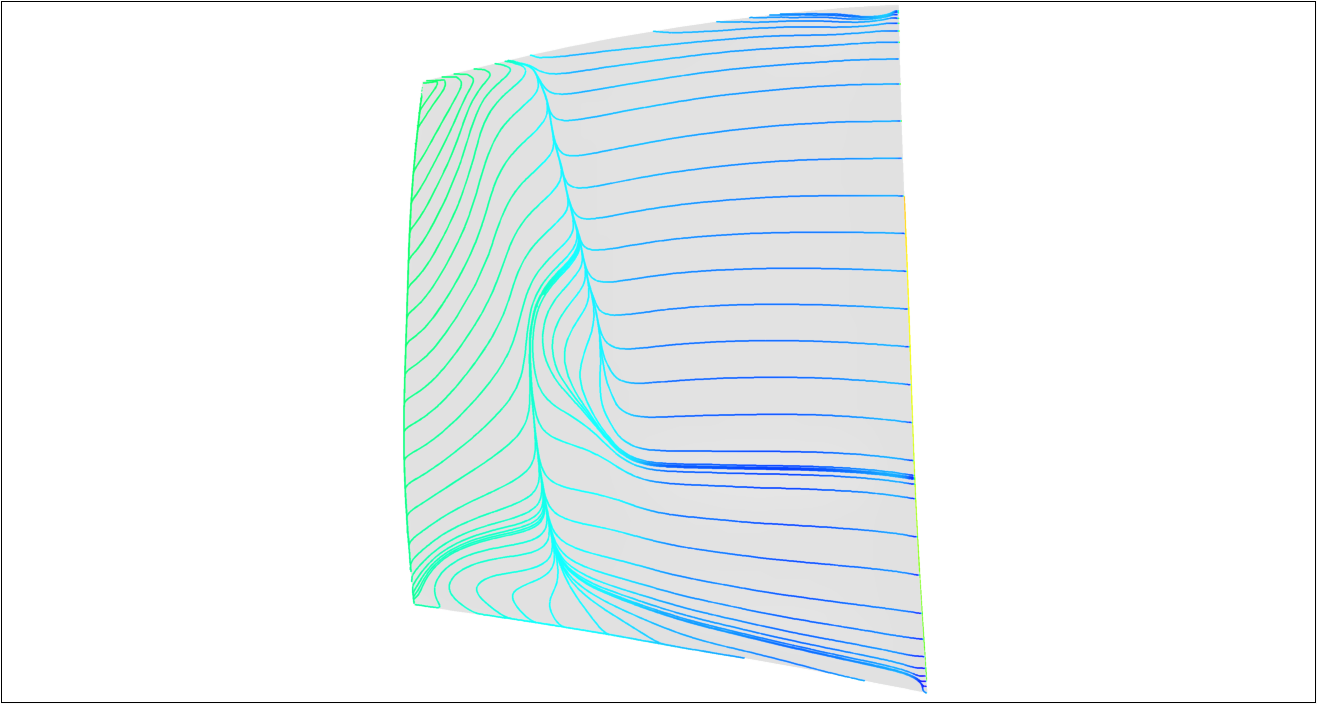}
\end{minipage}
}%
\subfigure[Nash-EGO]{
\begin{minipage}[t]{0.33\linewidth}
\centering
\includegraphics[scale=0.26,trim=11cm 0.5cm 11cm 0.2cm,clip]{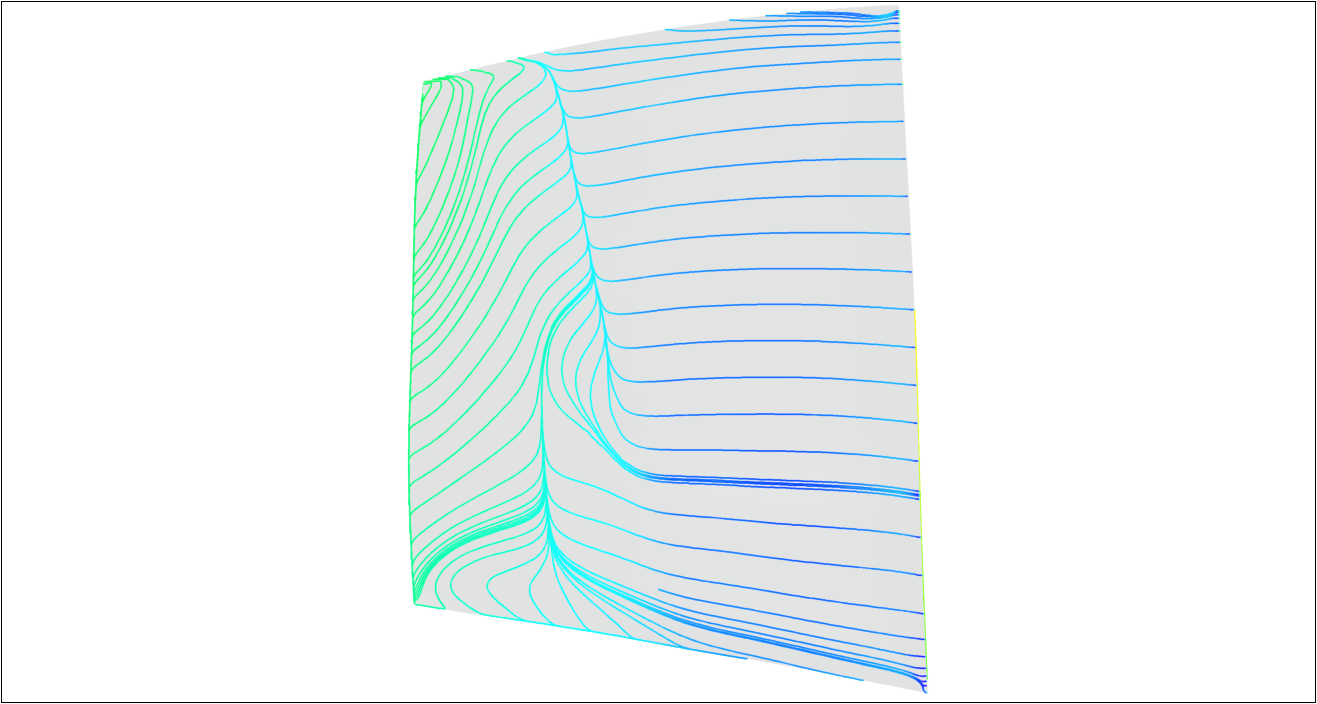}
\end{minipage}
}%

\subfigure[IKAEA]{
\begin{minipage}[t]{0.33\linewidth}
\centering
\includegraphics[scale=0.26,trim=11cm 0.5cm 11cm 0.2cm,clip]{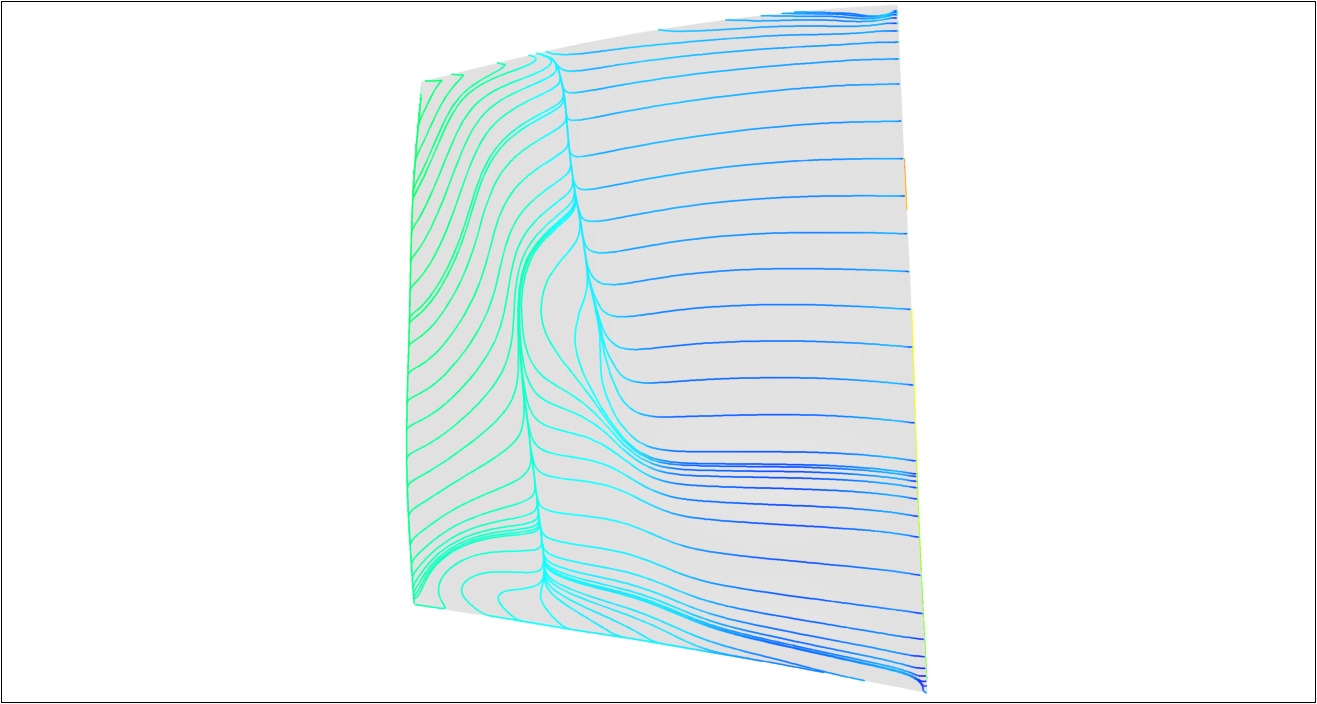}
\end{minipage}%
}%
\subfigure[GA]{
\begin{minipage}[t]{0.33\linewidth}
\centering
\includegraphics[scale=0.26,trim=11cm 0.5cm 11cm 0.2cm,clip]{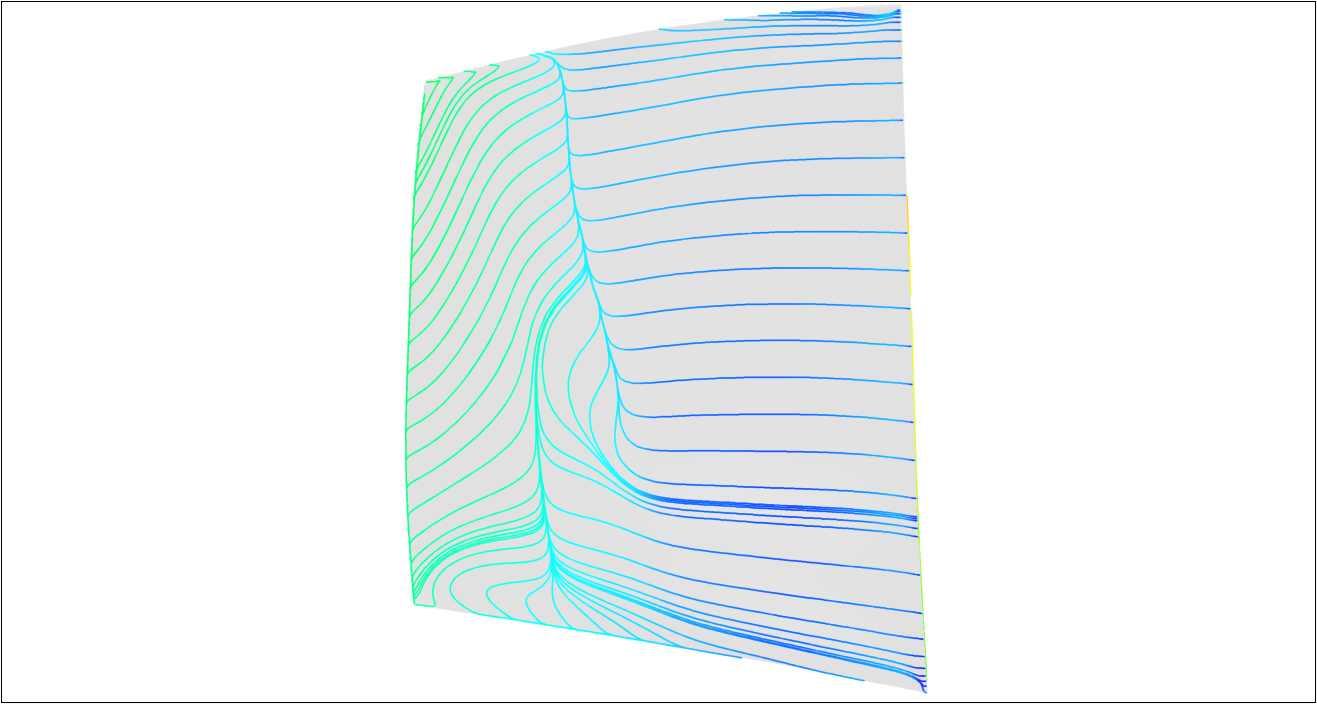}
\end{minipage}
}%
\subfigure[DE]{
\begin{minipage}[t]{0.33\linewidth}
\centering
\includegraphics[scale=0.26,trim=11cm 0.5cm 11cm 0.2cm,clip]{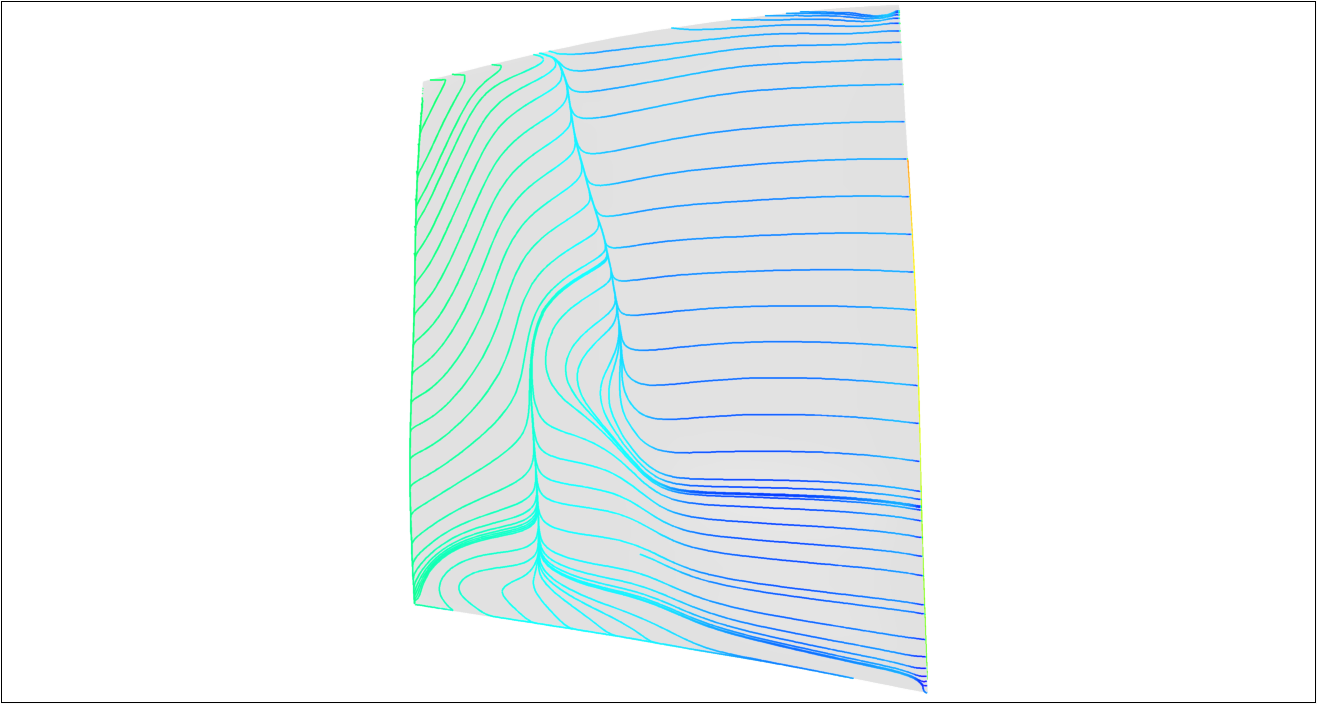}
\end{minipage}
}%
\centering
\caption{The limit streamlines the suction surface of the optimal designs with different algorithms and the baseline design.}
\label{fig:kt-original-12}
\end{figure}

In the 5 repeated optimization, the median optimal result obtained by each algorithm is selected for more specific analysis.
Figure~\ref{fig:kt-original-12} shows the suction surface's limit flow of the baseline design and the optimization results of different algorithms. 
It can be seen from Figure~\ref{fig:kt-original-12}(a) that there is a "shock wave-boundary layer interference" on the suction surface of the baseline Rotor 37 blade, and there is a serious flow separation phenomenon near the trailing edge of the suction surface. 
There is also an involved back-flow near the trailing edge of the tip. 
Compared with Figure~\ref{fig:kt-original-12}(b), the optimized separation line moves downstream and the separation area decreases, which is beneficial to the improvement of efficiency. 
And the intensity of the involved back-flows is also reduced.
\par
The above engineering results show that the KT-EGO algorithm is feasible and practical in large variable engineering optimization design, and it is also proved that the algorithm has obvious advantages in efficiency compared with other common methods.
Hence, the effectiveness of our proposed KT-EGO algorithm has been demonstrated.
\section*{5. Conclusions}
\par 

In this paper, we presented a knowledge transfer assisted efficient global optimization algorithm, namely KT-EGO, to solve the high-dimensional expensive black-box problems more efficiently.
Specifically, after dividing the high-dimensional design space into subset design spaces, we propose a surrogate-based data fusion strategy to gain knowledge from the previously evaluated samples to accelerate the optimization progress of the subset design space.
And further, at the later stage of the optimization process, we propose a strategy with adaptive variable range to enhance the local exploitation of the subset design space. 
Moreover, considering the joint effects of design variables, we propose a random decomposition strategy to better fit the landscape of the original black-box problem.
Through tests on 12 benchmark functions ranging from 30 dimensions to 60 dimensions and an engineering problem of compressor blade design, the effectiveness of our proposed algorithm has been well demonstrated.

\section*{Funding}
This work was supported by the National Science and Technology Major Project (2019-II-0008-0028) and the National Science Foundation of China (No. 51936008), and HFZL2021CXY004.
\section*{Acknowledgments}
The authors would like to thank the anonymous referees for their valuable comments.
\section*{Conflict of interest}
The authors declare that they have no conflict of interest.
\section*{Data availability statement}
The data that support the findings of this study are available from the corresponding author upon reasonable request.
The published article identifies the code repository as \url{https://github.com/zhet1997/KT-EGO_publish}; current repository availability is not asserted here.

\bibliographystyle{tfcad}
\bibliography{interactcadsample}%

\end{document}